\documentstyle[draft,psfig]{mn} 

\newif\ifAMStwofonts
\ifoldfss
  \ifCUPmtlplainloaded \else
    \NewTextAlphabet{textbfit} {cmbxti10} {}
    \NewTextAlphabet{textbfss} {cmssbx10} {}
    \NewMathAlphabet{mathbfit} {cmbxti10} {} % for math mode
    \NewMathAlphabet{mathbfss} {cmssbx10} {} %  "   "    "
  \fi
  \ifAMStwofonts
    \ifCUPmtlplainloaded \else
      \NewSymbolFont{upmath} {eurm10}
      \NewSymbolFont{AMSa} {msam10}
      \NewMathSymbol{\upi}     {0}{upmath}{19}
      \NewMathSymbol{\umu}     {0}{upmath}{16}
      \NewMathSymbol{\upartial}{0}{upmath}{40}
      \NewMathSymbol{\leqslant}{3}{AMSa}{36}
      \NewMathSymbol{\geqslant}{3}{AMSa}{3E}

      \let\leq=\leqslant 
      \let\geq=\geqslant 
    \fi
  \fi
\fi % End of OFSS

\ifnfssone
  \newmathalphabet{\mathit}
  \addtoversion{normal}{\mathit}{cmr}{m}{it}
  \addtoversion{bold}{\mathit}{cmr}{bx}{it}
  \newmathalphabet{\mathbfit} % math mode version of \textbfit{..}
  \addtoversion{normal}{\mathbfit}{cmr}{bx}{it}
  \addtoversion{bold}{\mathbfit}{cmr}{bx}{it}
  \newmathalphabet{\mathbfss} % math mode version of \textbfss{..}
  \addtoversion{normal}{\mathbfss}{cmss}{bx}{n}
  \addtoversion{bold}{\mathbfss}{cmss}{bx}{n}
  \ifAMStwofonts
    \ifCUPmtlplainloaded \else
      \UseAMStwoboldmath
      \makeatletter
      \new@mathgroup\upmath@group
      \define@mathgroup\mv@normal\upmath@group{eur}{m}{n}
      \define@mathgroup\mv@bold\upmath@group{eur}{b}{n}
      \edef\UPM{\hexnumber\upmath@group}
      \new@mathgroup\amsa@group
      \define@mathgroup\mv@normal\amsa@group{msa}{m}{n}
      \define@mathgroup\mv@bold\amsa@group{msa}{m}{n}
      \edef\AMSa{\hexnumber\amsa@group}
      \makeatother
      \mathchardef\upi="0\UPM19
      \mathchardef\umu="0\UPM16
      \mathchardef\upartial="0\UPM40
      \mathchardef\leqslant="3\AMSa36
      \mathchardef\geqslant="3\AMSa3E

      \let\leq=\leqslant 
      \let\geq=\geqslant 
    \fi
  \fi
\fi % End of NFSS release 1

\ifnfsstwo
  \DeclareMathAlphabet{\mathbfit}{OT1}{cmr}{bx}{it}
  \SetMathAlphabet\mathbfit{bold}{OT1}{cmr}{bx}{it}
  \DeclareMathAlphabet{\mathbfss}{OT1}{cmss}{bx}{n}
  \SetMathAlphabet\mathbfss{bold}{OT1}{cmss}{bx}{n}
  \ifAMStwofonts
    \ifCUPmtlplainloaded \else
      \DeclareSymbolFont{UPM}{U}{eur}{m}{n}
      \SetSymbolFont{UPM}{bold}{U}{eur}{b}{n}
      \DeclareSymbolFont{AMSa}{U}{msa}{m}{n}
      \DeclareMathSymbol{\upi}{0}{UPM}{"19}
      \DeclareMathSymbol{\umu}{0}{UPM}{"16}
      \DeclareMathSymbol{\upartial}{0}{UPM}{"40}
      \DeclareMathSymbol{\leqslant}{3}{AMSa}{"36}
      \DeclareMathSymbol{\geqslant}{3}{AMSa}{"3E}

      \let\leq=\leqslant 
      \let\geq=\geqslant 
    \fi
  \fi
\fi % End of NFSS release 2

\ifCUPmtlplainloaded \else
  \ifAMStwofonts \else % If no AMS fonts
    \def\upi{\pi}
    \def\umu{\mu}
    \def\upartial{\partial}
  \fi
\fi

\title[Black hole--neutron star coalescence]{Newtonian hydrodynamics
of the coalescence of black holes with neutron stars III: Irrotational
binaries with a stiff equation of state}

\author[W. H. Lee]
{William H. Lee \\
Instituto de Astronom\'{\i}a, Universidad Nacional Aut\'{o}noma
de M\'{e}xico, Apdo. Postal 70--264, Cd. Universitaria, 04510
M\'{e}xico D.F.\\
}
\pagerange{\pageref{firstpage}--\pageref{lastpage}}
\pubyear{1999}

\begin{document}

\maketitle

\label{firstpage}

\begin{abstract}
We present a numerical study of the hydrodynamics in the final stages
of inspiral in a black hole--neutron star binary, when the binary
separation becomes comparable to the stellar radius. We use a
Newtonian three--dimensional Smooth Particle Hydrodynamics (SPH) code,
and model the neutron star with a stiff (adiabatic index $\Gamma=3$
and $\Gamma=2.5$) polytropic equation of state and the black hole as a
Newtonian point mass which accretes matter via an absorbing boundary
at the Schwarzschild radius. Our initial conditions correspond to
irrotational binaries in equilibrium (approximating the neutron star
as a compressible tri--axial ellipsoid), and we have explored
configurations with different values of the initial mass ratio
$q=M_{\rm NS}/M_{\rm BH}$, ranging from $q=0.5$ to $q=0.2$. The
dynamical evolution is followed using an ideal gas equation of state
for approximately 23~ms. We have included gravitational radiation
losses in the quadrupole approximation for a point--mass binary. For
the less compressible case ($\Gamma=3$), we find that after an initial
episode of intense mass transfer, the neutron star is not completely
disrupted and a remnant core remains in orbit about the black hole in
a stable binary configuration. For $\Gamma=2.5$---which is believed to
be appropriate for matter at nuclear densities---the tidal disruption
process is more complex, with the core of the neutron star surviving
the initial mass transfer episode but being totally disrupted during a
second periastron passage. The resulting accretion disc formed around
the black hole contains a few tenths of a solar mass. A nearly
baryon--free axis is present in the system throughout the coalescence,
and only modest beaming of a relativistic fireball that could give
rise to a gamma--ray burst would be sufficient to avoid excessive
baryon contamination. We find that some mass (on the order of
$10^{-2}$~M$_{\odot}$) may be dynamically ejected from the system, and
could thus contribute substantially to the amount of observed
r--process material in the galaxy. We calculate the gravitational
radiation waveforms and luminosity emitted during the coalescence in
the quadrupole approximation, and show that they directly reflect the
morphology of the coalescence process. Finally, we present the results
of dynamical simulations that have used spherical neutron stars
relaxed in isolation as initial conditions, in order to gauge the
effect of using non--equilibrium initial conditions on the evolution
of the system.
\end{abstract}


\section{Introduction and motivation}

The emission of gravitational waves in a binary system and the
accompanying loss of angular momentum produce a decrease in
separation, and for certain systems, will inevitably lead to
coalescence (if the initial separation is small enough so that the
decay will take place in less than the Hubble time). In fact, the
binary neutron star systems known---PSR~1913+16 (Hulse \&
Taylor~1975), PSR~1534+12 (Wolszczan~1991)---have been observed to
decay at a rate that matches the prediction of general relativity to
high accuracy (Taylor et al.~1992; Stairs et al.~1998). There are no
observed black hole--neutron star systems yet, but it is believed that
they do in fact exist. Estimates of the event rates can be inferred
from the statistics of Hulse--Taylor type systems and from theoretical
studies of stellar evolution, and are expected to be about 10$^{-6}$
to 10$^{-5}$ per year per galaxy, implying several coalescences per
year out to a distance of 1~Gpc could be taking place (Lattimer \&
Schramm~1976; Narayan, Piran \& Shemi~1991; Tutukov \& Yungelson~1993;
Lipunov, Postnov \& Prokhorov~1997; Portegies Zwart \& Yungelson~1998;
Bethe \& Brown~1998; Kalogera~1998; Be\l czy\'{n}ski \& Bulik~1999).

The astrophysical interest in compact binaries such as the ones
treated here is varied. They are primary candidates for detection by
the gravitational detectors being currently constructed and expected
to begin operation within the next few years, such as LIGO (Abramovici
et al.~1992) and VIRGO (Bradaschia et al.~1990). When the binary
separation is large (compared with the stellar radius), the system can
be effectively considered to consist of point masses, and accurate
waveforms can be calculated using the post newtonian approximation
(see e.g. Kidder, Will \& Wiseman~1992; Cutler et al.~1993; Blanchet
et al.~1995). Detection of these signals will yield information about
the masses and spins of the system. At small separations, a powerful
but short burst of waves containing information about the radii and
internal structure of the stars is expected. In particular, it may
help constrain the equation of state of matter at high densities. At
this stage, hydrodynamics will play an important role in the evolution
of the system and must be taken into account to make any sense of
possible observations. Clearly, this requires performing accurate,
three--dimensional simulations in general relativity. These
simulations have not yet been performed, but progress is being made in
this direction by several groups (see e.g. Wilson, Mathews \&
Marronetti~1996; Lombardi, Rasio \& Shapiro~1997; Baumgarte et
al.~1997; Shibata~1999; Shibata \& Ury\={u}~2000).

Theoretical studies of the tidal disruption and coalescence of black
holes with neutron stars were carried out many years ago
(Wheeler~1971; Lattimer \& Schramm~1974, 1976) but the computing
resources to perform accurate numerical studies have only recently
become available. Numerical work on the final stages of binary
evolution and close interactions, including coalescence, has been
carried out in the Newtonian case by many authors. These calculations
may serve as a useful guide for future work, and also provide
benchmarks one can use to compare the results of future general
relativistic codes when used in the weak--field limit. The first
studies of binary neutron star coalescence were done by Oohara and
Nakamura~(1989) who focused on the emission of gravitational
waves. They later improved their work, using equilibrium conditions
for tidally locked systems, and including the effects of radiation
reaction in their calculations (Nakamura \& Oohara~1989, 1991; Oohara
\& Nakamura~1990, 1992). Lai, Rasio \& Shapiro~(1993b, hereafter LRSb)
developed a semi--analytical method that allowed them to generalize
the study of incompressible rotating ellipsoids done by
Chandrasekhar~(1969) to compressible ones, using throughout a
polytropic equation of state. Rasio \& Shapiro~(1992, 1994, 1995,
hereafter RS92, RS94, RS95) extended this work to dynamical
simulations of coalescing binaries using Smooth Particle Hydrodynamics
(SPH), exploring the effect of the mass ratio and the stiffness of the
equation of state on the global behavior of the system, as well as on
the gravitational radiation signal. They clearly showed that
hydrodynamics can play a crucial role in the evolution of such
systems, making them dynamically unstable for small separations even
in the Newtonian regime (Lai, Rasio \& Shapiro 1993a, hereafter
LRSa). Zhuge, Centrella \& McMillan~(1994, 1996) also used SPH and
concentrated on the gravitational wave spectrum and on the effect of
varying the neutron star spin and radius. The thermodynamic evolution
of the fluid during and after coalescence has been studied using a
physical equation of state by Davies et al.~(1994) and Ruffert, Janka
\& Sch\"{a}fer~(1996) in double neutron star mergers, and recently by
Janka et al.~(1999) in black hole--neutron star mergers. This has
important implications for gamma ray burst models (see below).

It is now believed that the gamma ray bursts (GRBs) are of
cosmological origin (Meegan et al.~1992; see Fishman \& Meegan~1995
for a review), with redshifts to several optical afterglows
(M\'{e}sz\'{a}ros \& Rees~1997a) having been measured recently
(Metzger et al.~1997; Djorgovski et al.~1998; Kulkarni et al.~1998,
1999). The extreme energetics of these events (e.g. Kulkarni et
al.~1999) and their variability (which arises at the source, Sari \&
Piran~1997) indicates that the ``central engine'' must involve a
compact object of some sort. The preferred model involves the
relativistic expansion of a fireball that produces the gamma rays as a
result of internal shocks in the ejecta. Beaming of this fireball
would reduce the energy requirements somewhat, and recent observations
indicate that this could indeed be the case (Harrison et al.~1999;
Stanek et al.~1999). What the central engine itself is made of remains
to be seen. The bimodality in burst durations (Kouveliotou et
al.~1995) with classes of long and short bursts (separated at about
2~seconds) hint that there might be two different processes producing
the GRBs. Different mechanisms involving neutron stars and/or black
holes have been proposed, such as binary coalescence of neutron stars
(Paczy\'{n}ski~1986; Goodman~1986; Eichler et al.~1989; Narayan,
Paczy\'{n}ski \& Piran~1992), neutron stars with black holes
(Paczy\'{n}ski~1991), catastrophic release of rotational energy
through intense magnetic fields (Usov~1992; Klu\'{z}niak \&
Ruderman~1998; Spruit~1999) and failed supernova
explosions~(Woosley~1993). Janka \& Ruffert~(1996) estimated that the
energy release in neutrinos during a binary neutron star coalescence
was insufficient to power a GRB. However, after the initial
coalescence an accretion torus may be formed around a resulting black
hole, and such a structure might produce a GRB (Jarozy\'{n}ski~1993,
1996; Witt et al.~1994; Popham, Woosley \& Fryer~1999; MacFadyen \&
Woosley~1999). The fireball might be powered by neutrino emission from
this disc (e.g. Goodman, Dar \& Nussinov~1987; Thompson~1994; Ruffert
\& Janka~1999) or by the energy of rotation of the black hole at the
center of the disc (M\'{e}sz\'{a}ros \& Rees~1997b) if one is present
(Blandford \& Znajek~1977).

It is not clear that the r--process nuclei can be produced by
supernovae explosions alone in the observed abundances (see e.g. Meyer
\& Brown~1997; Freiburghaus et al.~1999a). Thus an alternative
mechanism seems desirable. The ejection of nuclear matter during
binary neutron star or black hole--neutron star mergers and its
subsequent decompression could provide an adequate environment for
this to occur, as has been suggested before (Lattimer \&
Schramm~1974,~1976; Symbalisty \& Schramm~1982; Eichler et
al.~1989). Recently, Newtonian hydrodynamical simulations by Rosswog
et al.~(1999) have addressed this problem, and the evolution of the
ejected matter has been studied in greater detail by Freiburghaus,
Rosswog \& Thielemann~(1999b). So far, they have found that indeed
binary neutron star mergers seem to be promising candidates for this
process to occur. Since the rates of black hole--neutron star mergers
are probably comparable (see above), it seems natural to assume that
this will also happen in such systems. Since our equation of state is
of a simple form, we do not calculate nuclear reactions in any way,
but merely wish to determine how much (if any) matter can be ejected
during the dynamical coalescence and/or tidal disruption of the
neutron star by the black hole, a necessary first step if it is to
contribute to the abundances of r--process material.

In previous work (Lee \& Klu\'{z}niak~1995; Klu\'{z}niak \& Lee~1998;
Lee \& Klu\'{z}niak~1999a,b---hereafter Papers I and II) we have
studied the dynamical coalescence of tidally locked black
hole--neutron star systems. In all cases the neutron star was modeled
as a polytrope and the dynamical simulation was followed using an
ideal gas equation of state, as we have done for this work. We studied
the case of a stiff ($\Gamma=3$, Paper~I) and a soft ($\Gamma=5/3$,
Paper~II) equation of state, mainly to gauge what effect this would
have on the outcome of the coalescence. We found that the
compressibility of the neutron star affected the process greatly, with
the binary surviving after a brief episode of mass transfer for the
former, and unstable mass transfer leading to complete tidal
disruption in the latter. We also confirmed through these dynamical
simulations the presence of a dynamical instability for high mass
ratios before Roche lobe overflow ocurred for $\Gamma=3$. In the case
of $\Gamma=5/3$ mass transfer was initiated before the onset of
dynamical instability (see Figure~1 in Paper~II). In all cases we
found that the outcome of the event was extremely favorable for the
production of gamma ray bursts, with a region of low density along the
rotation axis of the binary, and in some cases an accretion disc
containing a few tenths of a solar mass around the black hole. This
scenario seems particularly well suited for the class of short GRBs
(Klu\'{z}niak \& Lee~1998; see Fryer, Woosley \& Hartmann~1999 for a
variety of scenarios that could give rise to a GRB).

In the present paper, we extend our work to the case of irrotational
binaries, which is believed to be a more realistic approximation to
the conditions encountered in such systems (see below,
section~\ref{initial}). We have chosen to model the neutron star with
a stiff equation of state, using $\Gamma=3$ as before, but also
studying the case with $\Gamma=2.5$. This may seem like a small
difference, but the outcome is significantly altered, as shown
below. We have adopted this value because it closely matches the
adiabatic index that is believed to be appropriate for matter at
nuclear densities (e.g. in the physical equation of state of Lattimer
\& Swesty~1991).

The paper is organized as follows. In section~\ref{method} we present
the numerical method we have used, emphasizing the modifications we
have implemented in our code. The initial conditions used for the
dynamical simulations are presented in section~\ref{initial}, followed
by a presentation of the simulations themselves in
section~\ref{results}. The influence of the choice of initial
conditions is explored in section~\ref{sphere}. A summary and
discussion of our work is given in section~\ref{discussion}.

\section{Numerical method} \label{method}

For the calculations presented in this work, we have used the method
known as Smooth Particle Hydrodynamics (SPH) (see Monaghan~1992 for a
review and Lee~1998 for a description of our own code). The code is
essentially the same one that was used for our previous simulations of
tidally locked black hole--neutron star binaries (Paper~I, Paper~II)
with some minor modifications, discussed below. Here we will not
discuss the code in detail nor present any tests of the method, since
it has now become widely known. We limit the discussion of numerical
implementation to the new features of the code.

As before, the black hole is modeled as a Newtonian point mass of mass
$M_{BH}$ with an absorbing boundary at the Schwarzschild radius
$r_{Sch}=2GM_{\rm BH}/c^{2}$. Any SPH particle that crosses this
boundary is absorbed by the black hole, whose mass and momentum are
adjusted so as to ensure conservation of total mass and total linear
momentum in the system.

The neutron star is modeled as a polytrope with a stiff equation of
state, so that the pressure is given by $P=K \rho^{\Gamma}$ with
$\Gamma$ and $K$ being constants (see Paper~I). Unless otherwise
noted, we measure mass and distance in units of the mass and radius of
the unperturbed (spherical) neutron star (13.4~km and 1.4~M$_{\odot}$
respectively), so that the units of time, density and velocity are
\begin{eqnarray}
\tilde{t}=1.146\times 10^{-4}{\rm s}\times \left( \frac{R}{13.4~{\rm
km}}\right) ^{3/2} \left( \frac{M_{\rm NS}}{1.4~M_{\odot}}\right)
^{-1/2} \label{eq:deftunit}
\end{eqnarray}
\begin{eqnarray}
\tilde{\rho}=1.14\times 10^{18}{\rm kg~m^{-3}}\times \left(
\frac{R}{13.4~{\rm km}}\right) ^{-3} \left( \frac{M_{\rm
NS}}{1.4~M_{\odot}}\right) \label{eq:defrhounit}
\end{eqnarray}
\begin{eqnarray}
\tilde{v}=0.39 c \times \left( \frac{R}{13.4~{\rm km}}\right) ^{-1/2}
\left( \frac{M_{\rm NS}}{1.4~M_{\odot}}\right) ^{1/2} \label{eq:defvunit}
\end{eqnarray}
For the dynamical simulations presented here, we have used $N\simeq
38000$ SPH particles to model the neutron star (except where
noted). The initial (spherical) neutron star is constructed by placing
the SPH particles on a uniform three--dimensional grid with particle
masses proportional to the Lane--Emden density. This ensures that the
spatial resolution is approximately uniform throughout the fluid. This
isolated star is then allowed to relax for a period of thirty time
units (as defined above) by including a damping term linear in the
velocities in the equations of motion. The specific entropies of the
particles are kept constant during this procedure (i.e. {\em K} is
constant in the equation of state $P=K \rho^{\Gamma}$).

To perform a dynamical run, the black hole and every SPH particle are
given the velocity as determined from the corresponding initial
condition (see below) in an inertial frame, with the origin of
coordinates at the centre of mass of the system. Each SPH particle is
assigned a specific internal energy $u_{i}=K
\rho^{(\Gamma-1)}/(\Gamma-1)$ and the equation of state is changed to
that of an ideal gas, $P=(\Gamma-1)\rho u$. The specific internal
energy is then evolved according to the first law of thermodynamics,
taking into account the contribution from artifical viscosity (see
below). We vary the initial mass ratio $q=M_{\rm NS}/M_{\rm BH}$ in
the binary by adjusting the mass of the black hole only.

\subsection{Artificial viscosity}

Artificial viscosity is used in SPH to handle the presence of shocks
and avoid particle interpenetration. The momentum and energy equations
are written as:
\begin{eqnarray*}
\frac{d\mbox{\boldmath $v$}_{i}}{dt}=-\sum_{j}m_{j}\left(
2\frac{\sqrt{P_{i}P_{j}}}{\rho_{i}\rho_{j}}+\Pi_{ij} \right)
\mbox{\boldmath $\nabla$}_{i} W_{ij}-\mbox{\boldmath $\nabla$}
\Phi_{i}-\mbox{\boldmath $a$}^{\rm RR}_{i},
\end{eqnarray*}
and
\begin{eqnarray*}
\frac{du_{i}}{dt}=\frac{1}{2}\sum_{j}m_{j}\left(
2\frac{\sqrt{P_{i}P_{j}}}{\rho_{i}\rho_{j}}+\Pi_{ij} \right)
(\mbox{\boldmath $v$}_{i}-\mbox{\boldmath $v$}_{j}) \cdot
\mbox{\boldmath $\nabla$}_{i} W_{ij}.
\end{eqnarray*}
Here $\Pi_{ij}$ is the artificial viscosity term and $\mbox{\boldmath
$v$}$, $P$, $\rho$, $u$, $\Phi$, $\mbox{\boldmath $a$}^{\rm RR}_{i}$
and $W$ are the velocity, pressure, density, internal energy per unit
mass, gravitational potential, radiation reaction acceleration (see
below) and interpolation kernel respectively (for the kernel we use
the form of Monaghan \& Lattanzio (1985), see also Paper~I). In
previous work we have used the form introduced by Monaghan (1992),
containing both shear and bulk viscosity. We have now changed the
viscosity prescription and use the form presented by Balsara (1995),
namely:
\begin{eqnarray*}
\Pi_{ij}=\left( \frac{P_{i}}{\rho_{i}^{2}}+ \frac{P_{j}}{\rho_{j}^{2}}
\right)(-\alpha\mu_{ij}+\beta\mu^{2}_{ij})
\end{eqnarray*}
where
\begin{eqnarray*}
\mu_{ij}=\left \{ \begin{array}{ll} \frac{(\mbox{\boldmath
$v$}_{i}-\mbox{\boldmath $v$}_{j}) \cdot (\mbox{\boldmath
$r$}_{i}-\mbox{\boldmath $r$}_{j})}{h_{ij}(|\mbox{\boldmath
$r$}_{i}-\mbox{\boldmath $r$}_{j}|^{2}/h_{ij}^{2})+\eta^{2}}
\frac{f_{i}+f_{j}}{2c_{ij}} & (\mbox{\boldmath
$v$}_{i}-\mbox{\boldmath $v$}_{j}) \cdot (\mbox{\boldmath
$r$}_{i}-\mbox{\boldmath $r$}_{j}) <0 \\ 0 & (\mbox{\boldmath
$v$}_{i}-\mbox{\boldmath $v$}_{j}) \cdot (\mbox{\boldmath
$r$}_{i}-\mbox{\boldmath $r$}_{j}) \geq 0
\end{array} \right.
\end{eqnarray*}
and $f_{i}$ is the form function for particle {\em i} defined by
\begin{eqnarray*}
f_{i}=\frac{|\mbox{\boldmath $\nabla$} \cdot \mbox{\boldmath
$v$}|_{i}}{|\mbox{\boldmath $\nabla$} \cdot \mbox{\boldmath
$v$}|_{i}+|\mbox{\boldmath $\nabla$} \times \mbox{\boldmath
$v$}|_{i}+\eta'c_{i}/h_{i}}.
\end{eqnarray*}
The factor $\eta'\simeq 10^{-4}$ in the denominator prevents numerical
divergences. The sound speed at the location of particle {\em i} is
denoted by $c_{i}$, and $\alpha$ and $\beta$ are constants. The
divergence and curl of the velocity field are evaluated through
\begin{eqnarray*}
(\mbox{\boldmath $\nabla$} \cdot \mbox{\boldmath
$v$})_{i}=\frac{1}{\rho_{i}}\sum_{j}m_{j}(\mbox{\boldmath
$v$}_{j}-\mbox{\boldmath $v$}_{i})\cdot \mbox{\boldmath
$\nabla$}_{i}W_{ij}
\end{eqnarray*}
and
\begin{eqnarray*}
(\mbox{\boldmath $\nabla$} \times \mbox{\boldmath
$v$})_{i}=\frac{1}{\rho_{i}}\sum_{j}m_{j}(\mbox{\boldmath
$v$}_{j}-\mbox{\boldmath $v$}_{i})\times \mbox{\boldmath
$\nabla$}_{i}W_{ij}
\end{eqnarray*}
We use this new form of the viscosity because it vanishes in regions
of large vorticity, when $ \mbox{\boldmath $\nabla$} \times
\mbox{\boldmath $v$}\gg \mbox{\boldmath $\nabla$} \cdot
\mbox{\boldmath $v$}$. Since in our dynamical simulations accretion
structures are often formed (where strong shear is present), we wish
to limit the effect of artificial viscosity on the evolution of such
structures, while at the same time retain the ability to deal with
strong shocks (in regions of strong compression where $\mbox{\boldmath
$\nabla$} \cdot \mbox{\boldmath $v$} \gg \mbox{\boldmath $\nabla$}
\times \mbox{\boldmath $v$} $, the form function {\em f} approaches
unity). Lombardi et al.~(1999) have performed a study of the effects
of different artificial viscosity prescriptions in several problems
involving SPH, and they suggest using $\alpha \simeq \beta \simeq
\Gamma/2$, where $\Gamma$ is the adiabatic index. For all the
calculations presented here we have used these values for $\alpha$ and
$\beta$.

\subsection{Gravitational radiation reaction} \label{grr}

We have modified our implementation of gravitational radiation
reaction. Previously (see Paper~II, also Davies et al.~1994) we used
the quadrupole formula for angular momentum loss, and applied a back
reaction force to each binary component. This approach is only
reasonable as long as the neutron star is not tidally disrupted and
can be reasonably approximated as a point mass. Usually this radiation
reaction force is switched off when the binary separation (defined as
the distance between the black hole and the centre of mass of the
tidally deformed neutron star) decreases below a tidal disruption
radius, given by $r_{tidal} \simeq R(M_{\rm BH}/M_{\rm
NS})^{1/3}$. However, there are situations in which the initial
episode of mass transfer in the binary does not disrupt the neutron
star completely. Instead, a massive core survives while at the same
time an accretion disc forms around the black hole (see the results
for a tidally locked binary with a mass ratio of unity in Lee \&
Klu\'{z}niak~1995; Klu\'{z}niak \& Lee~1998 and Paper~I). In this case
one still has a stable binary system, composed of the black hole and
the surviving core, plus an accretion torus around the black
hole. Gravitational radiation losses will affect the evolution of the
new binary, but the system composed of the accretion torus and the
core (the fluid) can hardly be considered to be a point mass. We thus
realized that identifying the core was necessary in order to apply a
gravitational raditation reaction term to it and the black hole. This
needed to be done while the simulation was progressing, in order to
update the calculation of angular momentum losses as needed.

In all cases where it was apparent that this identification was a
necessity, the system had evolved through Roche lobe overflow into a
configuration consisting of a stable binary (a low--mass neutron star
around the black hole) and an accretion torus around the black hole
(see Figure~4 in Paper~I). It was apparent that the core consisted of
matter that was gravitationally self--bound, and that the fluid in the
accretion torus was in quasi Keplerian orbits around the black
hole. Thus identifying the core became an issue of finding where the
gravitationally self--bound matter was located. Our calculation of
gravitational forces in the code uses a hierarchical tree structure
(see Benz et al.~1990), and it simultaneously calculates what the
gravitational potential energy is for every particle,
$\Phi_{i}=\Phi_{i}^{\rm self}+\Phi_{i}^{\rm BH}$. The first term
arises from the contribution of all SPH particles, and the second from
the presence of the black hole, given simply by $\Phi_{i}^{\rm
BH}=-GM_{\rm BH}m_{i}/|\mbox{\boldmath $r$}_{\rm BH}-\mbox{\boldmath
$r$}_{i}|$. There is also kinetic energy in the fluid that needs to be
considered, and we take this into account by calculating the kinetic
energy of the SPH particles in the instantaneous frame of reference
that is {\em co--moving} with the centre of mass of the core (to
remove the influence of the kinetic energy of orbital motion). Since
we do not know which SPH particles will make up the core beforehand,
we use the position of the centre of mass of the core from the
previous time step in the computation. Since in our initial condition
(see below) the star has not overflowed its Roche lobe, the centre of
mass of the core at $t=0$ coincides with the centre of mass of the
entire neutron star. Thus the kinetic energy we calculate is
$K_{i}=m_{i}(\delta v)^{2}/2$, where $\delta \mbox{\boldmath
$v$}=\mbox{\boldmath $v$}_{i}-\mbox{\boldmath $v$}_{\rm core}$. If the
total mechanical energy of an SPH particle, $E_{i}=\Phi_{i}^{\rm
self}+K_{i}$ is negative, then that particle is identified as
belonging to the core of the star. We note that in the simulations
presented here, we have not considered the internal energy of the
fluid for this determination (we have performed several tests
including this term, and the effect it has on the global evolution of
the system is negligible). We keep track of the total core mass
$M_{\rm core}$ and the position of its centre of mass. The radiation
reaction formula is then applied to the black hole and to every SPH
particle in the core as in Paper~II, producing an acceleration
\begin{eqnarray*}
\mbox{\boldmath $a$}^{i}=-\frac{1}{q(M_{\rm BH}+M_{\rm core})}
\frac{dE}{dt} \frac{\mbox{\boldmath $v$}^{\rm core}_{cm}}{(v^{\rm
core}_{cm})^{2}},
\end{eqnarray*}
while for the black hole the acceleration is given by
\begin{eqnarray*}
\mbox{\boldmath $a$}^{\rm BH}=-\frac{q}{M_{\rm BH}+M_{\rm core}}
\frac{dE}{dt} \frac{\mbox{\boldmath $v$}^{\rm BH}}{(v^{\rm BH})^{2}}.
\end{eqnarray*}
The energy loss is given by the quadrupole formula (Landau \&
Lifshitz~1975):
\begin{eqnarray*}
\frac{dE}{dt}=-\frac{32}{5} \frac{G^{4}(M_{\rm core}+M_{\rm
BH})(M_{\rm core}M_{\rm BH})^{2}}{(c|\mbox{\boldmath $r$}_{\rm
BH}-\mbox{\boldmath $r$}_{\rm core}|)^{5}}.
\end{eqnarray*}
The emission of gravitational radiation is calculated as described in
Papers~I and II (see also Finn~1989; Rasio \& Shapiro~1992).

Of course, even with this new formulation of gravitational radiation
reaction, the star can in principle be completely disrupted, so
throughout our dynamical calculations, we monitor the separation of
the black hole and the core, and compare it to the current tidal
disruption radius. For $\Gamma=3$ we switch gravitational radiation
off if the distance between the centre of mass of the core and the
black hole drops below $r_{tidal}$ (this never actually happened for
the simulations presented here), and for $\Gamma=2.5$ we do so when
(and if) the mass of the core drops below one tenth of the initial
neutron star mass (i.e. 0.14~M$_{\odot}$). We then perform an {\em a
posteriori} check to ensure that the gravitational radiation reaction
terms are not switched off too late or too early in the simulation. We
have altered the criterion used to turn off radiation reaction from
that used previously (Paper~II) because we noticed that there
sometimes occurred a slight jump in the slope of total angular
momentum as a function of time ($dJ/dt$) at the moment when radiation
reaction was turned off. The new procedure eliminates this
discontinuity, and when radiation reaction is turned off it is no
longer an important factor in the evolution of the system. To make
sure that this is so, we continuously compute the radiation reaction
timescale $t_{RR}^{-1}=256G^{3}M_{\rm BH}M_{\rm core}(M_{\rm
BH}+M_{\rm core})/(5r^{4}c^{5})$ and an estimate of the current
orbital period $t_{orb}=2\pi / \sqrt{G(M_{\rm BH}+M_{\rm
core})/r^{3}}$, where {\em r} is the separation between the black hole
and the centre of mass of the core. For the typical separations and
masses in the black hole--core binary, by the time the core mass has
dropped to 0.14~M$_{\odot}$, the radiation reaction timescale is much
longer (by at least an order of magnitude) than the current orbital
period.

\section{Initial conditions} \label{initial}

In previous work (Papers~I and II) we have used tidally locked binary
systems as initial conditions. This was done initially for two main
reasons. First, it is relatively easy to set up self--consistent
initial conditions for systems in a state of rigid rotation. Since we
are interested in equilibrium configurations, the system can be
relaxed in the co--rotating frame of the binary and Coriolis forces
can be ignored. Second, we wanted to perform accurate comparisons to
the results for two neutron stars using essentially the same approach
(Newtonian hydrodynamics and a polytropic equation of state) as RS92,
RS94, RS95. On the other hand, it has been known for some time that
realistically, tidal locking is not expected in such binary systems,
because the viscosity inside neutron stars is not large enough
(Bildsten \& Cutler 1992; Kochanek 1992).

Thus the case of irrotational systems is physically more
interesting. Unfortunately, it is also more complicated. In this
scenario the {\em shape} of the star is fixed in the co--rotating
frame, but there are internal motions with zero circulation. Each
component appears to be counter--spinning with the orbital angular
velocity, and in an inertial frame, the star has effectively no
spin. (We note here that, as pointed out by Rasio \& Shapiro~(1999),
for double neutron star systems this creates a vortex sheet at the
interface of the stars when they come into contact that is extremely
difficult to handle numerically. For black hole--neutron star systems
like the ones presented here, this problem does not exist, since there
is no contact surface between the neutron star and the black hole, but
a vacuum boundary condition.)  Constructing equilibrium initial
conditions for irrotational systems in black hole--neutron star
binaries is a problem that has only recently been addressed (Ury\={u}
\& Eriguchi 1999). A method for constructing approximate solutions was
developped by LRSb. They used an energy variational method to find
equilibrium solutions for a variety of binary systems assuming a
polytropic equation of state and approximating the stars as
compressible tri--axial ellipsoids. The black hole--neutron star
binaries we treat in this work correspond to irrotational
Roche--Riemann binaries (see Section~8 in LRSb). We have used this
approach in order to construct initial conditions for all the
dynamical runs presented here, except two (see below).

\begin{figure}
\psfig{width=7.5cm,file=jra.eps.a,angle=0,clip=}
\end{figure}
\begin{figure}
\psfig{width=7.5cm,file=jrb.eps.a,angle=0,clip=}
\end{figure}
\setcounter{figure}{0}
\begin{figure}
\psfig{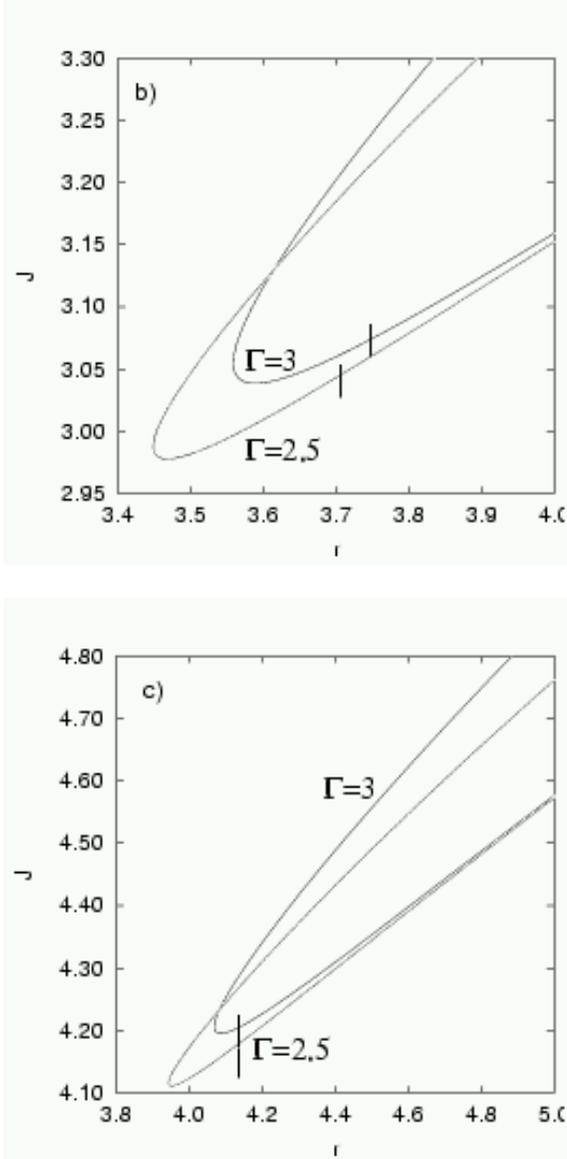}
\caption{Total angular momentum as a function of binary separation for
irrotational Roche--Riemann binaries as calculated using the method of
LRSb for $\Gamma=3$ (solid lines) and $\Gamma=2.5$ (dashed lines) for
(a) $q=0.5$; (b) $q=0.31$; (c) $q=0.2$. The thick vertical lines mark
the values of the initial separations used for the dynamical
simulations.}
\label{jra}
\end{figure}

We build an initial condition by first constructing a spherical star
of given radius and mass, as described in section~2.  We then use the
method of LRSb to calculate the orbital angular velocity of the binary
and the semimajor axes of the Roche--Riemann ellipsoid for the
appropriate choice of adiabatic index, initial mass ratio and binary
separation (see Table~\ref{parameters}). The semimajor axes of the
fluid configuration can also be calculated from the SPH numerical
solution using
\begin{eqnarray*}
a_{i}=\sqrt{\frac{5 I_{ii}}{\kappa_{n}M_{\rm NS}}}
\end{eqnarray*}
where 
\begin{eqnarray*}
I_{ii}=\sum_{j}m_{j}(x^{i}_{j})^{2}.
\end{eqnarray*}
The stiffness of the equation of state enters these equations through
the parameter $\kappa_{n}$ ($\kappa_{n}=0.815$ for $\Gamma=3$ and
$\kappa_{n}=0.75$ for $\Gamma=2.5$). The first and second semimajor
axes of the tri--axial ellipsoid lie in the orbital plane, with the
first one along the line joining the two binary components. The third
axis is oriented perpendicular to this plane (along the axis of
rotation).  The position of each SPH particle is then re--scaled
(independently along each coordinate axis) so that the new fluid
configuration has the appropriate semimajor axes. It is this new
ellipsoid that is then used as an initial condition for the
corresponding dynamical run. The initial velocity is given by the
orbital angular velocity (for the azimuthal component) plus the
(small) radial velocity corresponding to point--mass inspiral. We show
in Figure~\ref{jra} the variation in total angular momentum as a
function of binary separation for irrotational Roche--Riemann binaries
(with various mass ratios and adiabatic indices in the equation of
state), as calculated using the method of LRSb. The curves exhibit a
turning point as the separation is decreased, indicating the presence
of an instability in the system. Strictly speaking, the ellipsoidal
approximation breaks down close to this turning point, and a full
equilibrium solution is necessary (see Ury\={u} \&
Eriguchi~1999). However, we have chosen the values of the initial
separation for our dynamical runs $r_{i}$ to be slightly {\em above}
the turning point (see Table~\ref{parameters}). The ellipsoidal
approximation is then still reasonable, and our equilibrium
configurations have not yet reached the point where the neutron star
will overflow its Roche lobe. When the dynamical simulation is
initiated, the separation will decrease due to the emission of
gravitational waves, and mass transfer will start promptly. If one
were to construct the appropriate Roche--Riemann ellipsoid at a large
separation and then allow the system to evolve dynamically through
gravitational wave emission, the neutron star would be spun--up to
some degree for several reasons. First, since the neutron star cannot
react instantly to changes in the tidal potential produced by the
black hole, a tidal lag angle will be induced as the separation
decreases, so that the tri--axial ellipsoid and the binary axis are no
longer aligned (see Lai, Rasio \& Shapiro~1994). This tidal lag
increases as the separation is decreased. Immediately before Roche
lobe overflow occurs, the presence of a dynamical instability can make
it quite large (on the order of $10^{\circ}$). Second, any amount of
viscosity will transfer orbital angular momentum to the spin of the
neutron star, contributing to orbital decay. This applies both to a
real, physical viscosity, but also to any numerical effect that might
be present in our scheme. We have thus tried to minimize the effect of
numerical noise and viscosity on the initial evolution of the binary
by proceeding as described above, always keeping in mind that our
initial condition is not fully self--consistent. We hope to improve on
this in future work. The initial separations we have chosen are
similar to what we have presented before for the case of tidally
locked black hole--neutron star binaries (Paper~I, Paper~II).

We also include in Table~\ref{parameters} the initial parameters for
two runs (A31S and B31S) that used initially spherical neutron stars
for the dynamical calculations, with a Keplerian orbital angular
velocity. We have performed these runs to gauge the effect
non--equilibrium initial conditions will have on the evolution of the
system. The results of these runs are presented in
section~\ref{sphere}.

\begin{table*}
 \caption{Basic parameters for each run.}
 \label{parameters}
 \begin{tabular}{@{}lccccccccc}
  Run & $q$ & $\Gamma$ & $r_{i}$ & $a_{2}/a_{1}$ & $a_{3}/a_{1}$  
        & $\tilde{\Omega}$  
        & $t_{rad}$
        & $t_{f}$ & $N$ \\
  A50   & 0.50 & 3.0 & 3.25 & 0.728 & 0.761 & 0.30016 & 200.0 & 200.0 
        & 38,352 \\
  A31   & 0.31 & 3.0 & 3.76 & 0.730 & 0.760 & 0.28518 & 200.0 & 200.0 
        & 38,352 \\
  A31S  & 0.31 & 3.0 & 3.76 & 1.000 & 1.000 & 0.28196 & 200.0 & 200.0 
        & 38,352 \\
  A20   & 0.20 & 3.0 & 4.15 & 0.654 & 0.691 & 0.29120 & 200.0 & 200.0 
        & 38,352 \\
  B50   & 0.50 & 2.5 & 3.25 & 0.780 & 0.805 & 0.29876 & 36.29 & 200.0 
        & 37,752 \\
  B31   & 0.31 & 2.5 & 3.70 & 0.761 & 0.786 & 0.29154 & 41.51 & 200.0 
        & 37,752 \\
  B31S  & 0.31 & 2.5 & 3.70 & 1.000 & 1.000 & 0.28884 & 38.56 & 200.0 
        & 37,752 \\
  B20   & 0.20 & 2.5 & 4.15 & 0.729 & 0.757 & 0.29229 & 33.95 & 200.0 
        & 37,752 \\

 \end{tabular}

 \medskip

The table lists for each run (labeled) the initial mass ratio, the
adiabatic index used, the initial orbital separation, the axis ratios
for the tri--axial ellipsoid used as an initial condition, the initial
orbital angular velocity of the binary, the time at which
gravitational radiation reaction is switched off in the simulation,
the time at which the simulation was stopped, and the initial number
of particles. The runs labeled A31S and B31S used an initially
spherical neutron star (otherwise irrotational Roche--Riemann
ellipsoids were used, see text for details).

\end{table*}

\section{Results} \label{results}

We now present the result of dynamical calculations performed using
the initial conditions described in section~\ref{initial}.

\subsection{Morphology of the mergers}

The binary separation decreases as a result of angular momentum losses
to gravitational radiation, and in every case the neutron star
overflows its Roche lobe within one orbital period, initiating mass
transfer to the black hole.
\begin{figure*}
\psfig{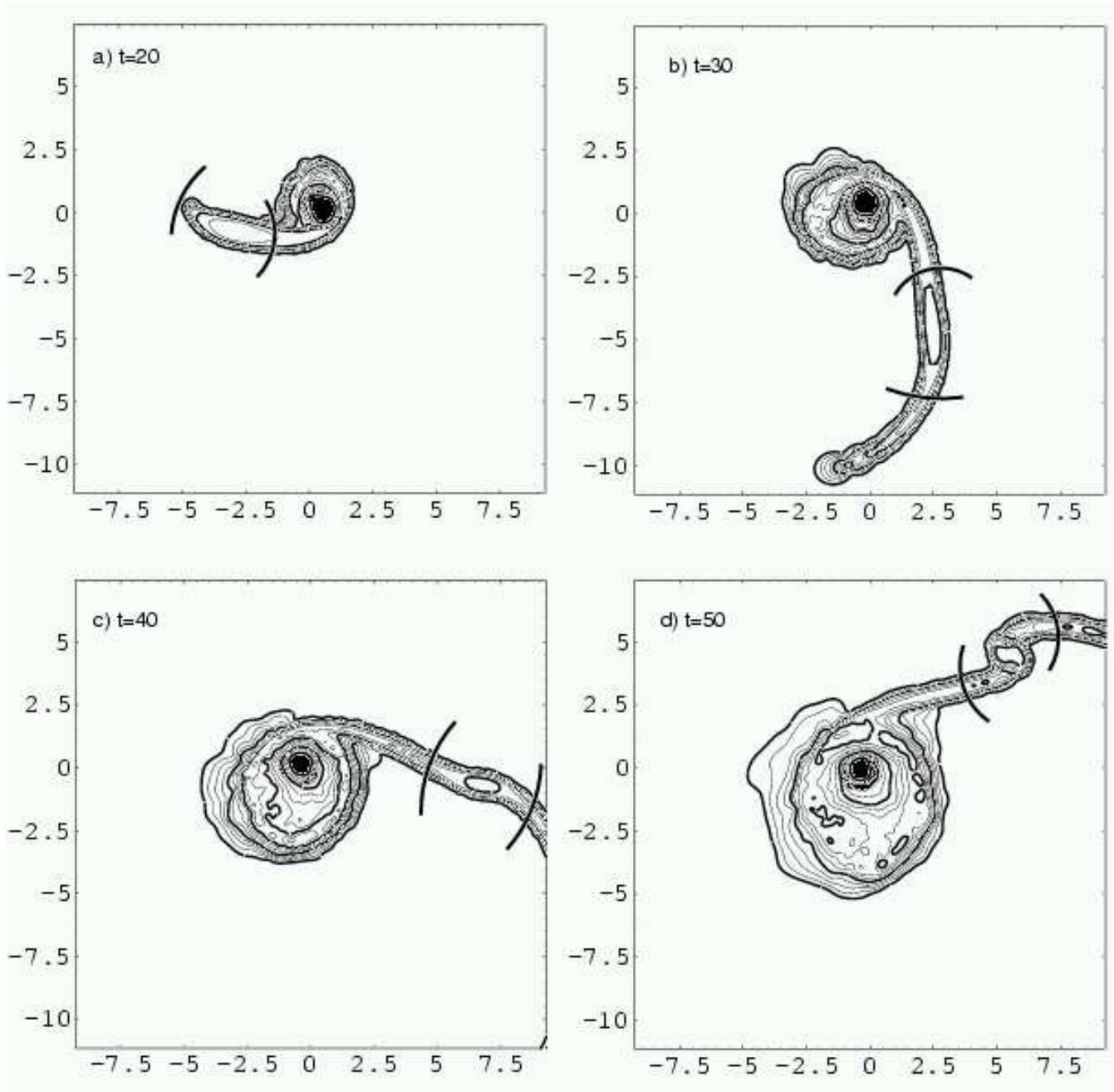}
\caption{Density contours in the orbital plane during the dynamical
simulation of the black hole neutron star binary with initial mass
ration $q=0.5$ and $\Gamma=3$ (run A50). The orbital rotation is
counterclockwise. All contours are logarithmic and equally spaced
every 0.25 dex. Bold contours are plotted at $\log \rho=-4,-3,-2,-1$
(if present) in the units defined in eq.~\ref{eq:defrhounit}. The
thick black arcs bound the matter that forms the core, according to
the procedure established in section~\ref{grr}. The time for each
frame is given in the top left corner, in the units defined in
eq.~\ref{eq:deftunit}.}
\label{rhog3}
\end{figure*}
\setcounter{figure}{1}
\begin{figure*}
\psfig{width=\textwidth,file=rhog3f2.eps.a,angle=0,clip=}
\caption{continued.}
\end{figure*}
\begin{figure*}
\psfig{width=\textwidth,file=rhog25f1.eps.a,angle=0,clip=}
\caption{Same as Figure~\ref{rhog3} but for run B31.}
\label{rhog25}
\end{figure*}
\setcounter{figure}{2}
\begin{figure*}
\psfig{width=\textwidth,file=rhog25f2.eps.a,angle=0,clip=}
\caption{continued.}
\end{figure*}
Figures~\ref{rhog3} and~\ref{rhog25} show a sequence of density
contour plots in the orbital plane of the binary for runs A50 and
B31. The neutron star becomes elongated (a), and an accretion stream
forms between it and the black hole. This stream winds around the
black hole, colliding with itself and producing a thick accretion
torus (b), (c), (d). At the same time as the accretion torus is being
formed, a long tidal tail of neutron star matter being ejected through
the outer Lagrange point appears. The most striking difference between
the two cases is that for $\Gamma=3$ (run A50) the neutron star core
clearly survives this first mass transfer episode as a coherent body,
while for $\Gamma=2.5$ (run B31) the star is almost completely
disrupted, although one can discern a bulge in the tidal tail in
panels (c) and (d) in Figure~\ref{rhog25}. The core (in the case of
run A50) and the bulge (in the case of run B31) make a second
periastron passage around the black hole--see panels (f,g,h). In the
first case, a second, less pronounced stream forms, feeding the
accretion disc, as well as a smaller, secondary tidal tail
(Figure~\ref{rhog3}g-h). In the latter case, the final disruption of
the core is not so evident, but one can see that the accretion disc
has a complex structure, with two partial rings on one side of the
disc (Figure~\ref{rhog25}g-h).
\begin{figure*}
\psfig{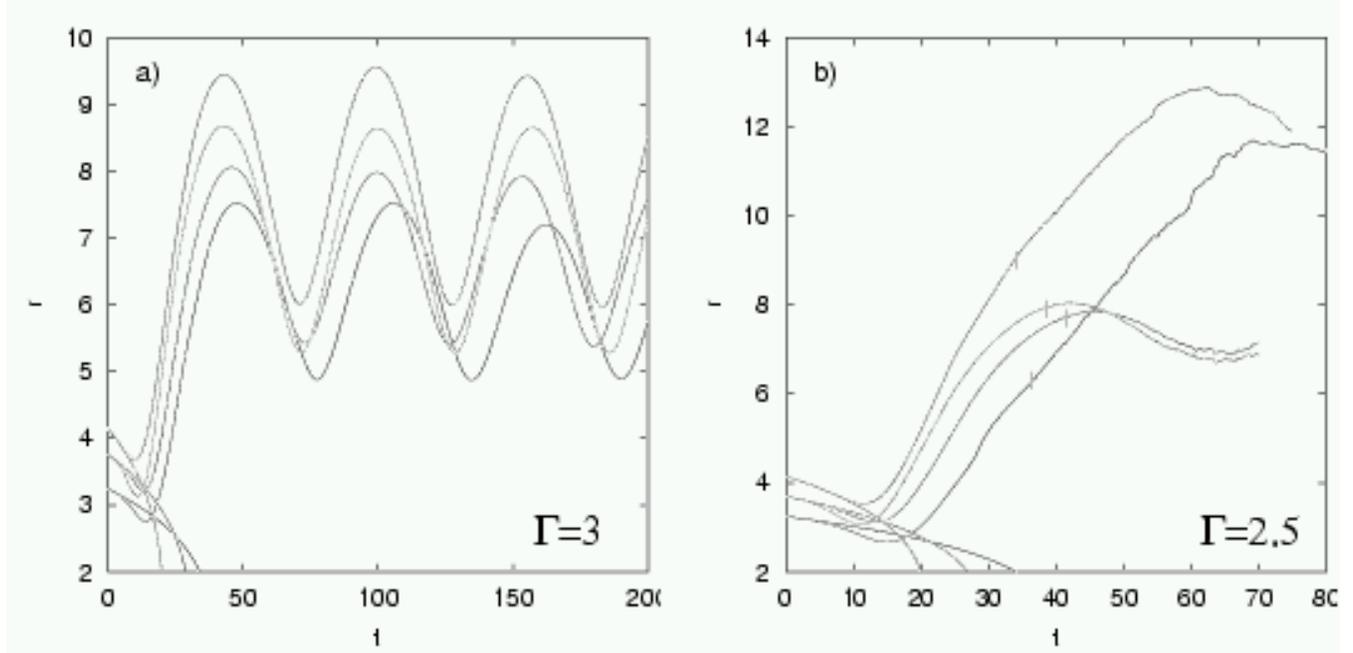}
\caption{Separation between the black hole and the centre of mass of
the core (as defined in section~\ref{grr}) as a function of time for
(a) $\Gamma=3$ and (b) $\Gamma=2.5$ (runs A50 and B50---solid lines;
runs A31 and B31---long--dashed lines; runs A31S and
B31S---short--dashed lines; runs A20 and B20--dotted lines). The
survival of the binary is apparent for the case with $\Gamma=3$. For
$\Gamma=2.5$ the curves terminate when the core is disrupted, close to
the second periastron passage, and a vertical line marks the time at
which the core mass drops below 0.14~M$_{\odot}$ and gravitational
radiation reaction is switched off. For $q=0.31$ there are two curves
in each frame, corresponding to runs initiated with a spherical
polytrope and an irrotational Roche--Riemann ellipsoid. In both cases,
the one that decays faster corresponds to the former condition. The
monotonically decaying curves correspond to point--mass binaries with
the same initial mass ratio and separation decaying through
gravitational wave emission, computed in the quadrupole
approximation.}
\label{rt}
\end{figure*}
The separation between the black hole and the centre of mass of the
core can be seen in Figure~\ref{rt}, where we plot it for all
runs. The separation initially decreases at a rate consistent with the
quadrupole approximation for a point--mass binary, and subsequently
deviates from this as the separation decreases at an even faster rate,
particularly for the higher mass ratios ($q=0.5$ and $q=0.31$). This
is a reflection of the importance of the hydrodynamics in the
evolution of the system. It was pointed out by LRSa that for such
compact binaries (black hole--neutron star or double neutron star
systems), a purely Newtonian dynamical instability can make the
orbital separation decay on an orbital timescale in the absence of
gravitational radiation losses, especially for relatively stiff
equations of state such as the ones being considered here. This is
precisely what is seen directly in the curves in Figure~\ref{rt}, and
also observed for the case of tidally locked binaries with a stiff
equation of state~(Paper~I).  The initial episode of mass transfer
then makes the separation increase, as we have seen above. In all
cases with $\Gamma=3$ the binary is not disrupted, and the surviving
core makes successive periastron passages, transferring some mass to
the accretion disc or directly to the black hole each time. We note
here that for all of these runs, the radiation reaction force is
always included in the equations of motion (see
Table~\ref{parameters}). For $\Gamma=2.5$, the neutron star is
completely disrupted after (at most) the second periastron passage,
and the accretion disc evolves steadily towards an axisymmetric
configuration. Our formulation of radiation reaction is strictly valid
only for circular orbits. However, the orbital eccentricity (when the
core survives) is small ($e\leq 0.2$) and thus we believe the errors
in the rate of energy and angular momentum loss are not excessive
(20\% and 10\% respectively). In all cases, the primary tidal tail
(formed during the first episode of mass transfer) persists as a
well--defined large scale structure throughout the simulation. At late
times, we see the formation of knots through a sausage instability at
approximately regular intervals all along the tail (this has been
observed in this kind of simulation before, see e.g.~RS94). Their
masses are fairly uniform, of order $2\times 10^{-3}$. This does not
occur for soft equations of state (with $\Gamma \leq 2$, see Paper~II
and RS92, RS95).

For a fixed value of the adiabatic index, varying the initial mass
ratio $q$ leads to qualitatively similar results, although there are
quantitative differences. The most important of these is that for
$\Gamma=3$ and lower mass ratios (runs A31 and A20) the amount of
matter that is left in orbit around the black hole after each
periastron passage is much smaller than for run A50, since most of it
is directly accreted by the black hole. This is in part due to the
fact that the black hole is physically larger in those cases, and thus
the infalling gas is more likely to cross the horizon than to wind
around the black hole and produce a larger accretion disc.

\begin{figure*}
\psfig{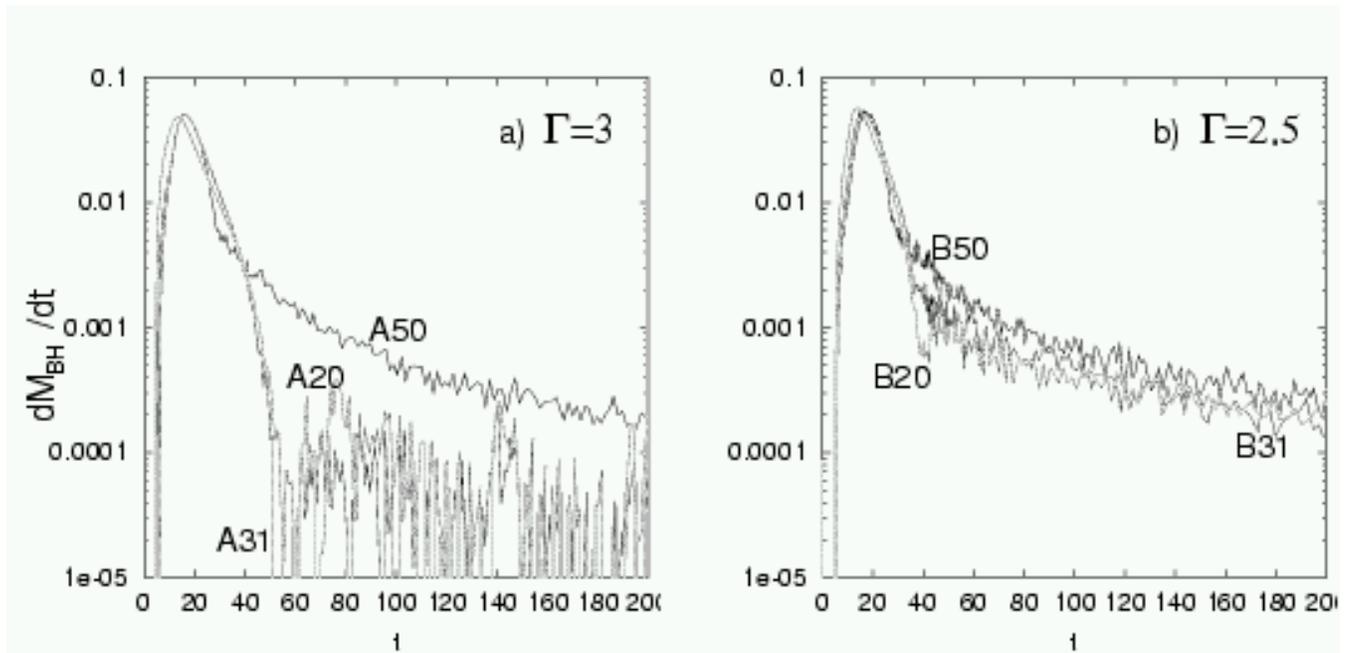}
\caption{Mass accretion rate onto the black hole as a function of time
for (a) $\Gamma=3$ and (b) $\Gamma=2.5$.}
\label{mdot}
\end{figure*}

We show in Figure~\ref{mdot} the mass accretion rates onto the black
hole as a function of time. The peak accretion rates (see
Table~\ref{disks}) occur during the initial episode of mass transfer,
and are of order 0.05 (equivalent to 0.6~M$_{\odot}$~ms$^{-1}$),
largely independent of the initial mass ratio and the adiabatic
index. There are however, important qualitative and quantitative
differences at later times. Whenever a torus forms around the black
hole, the accretion rate decreases rather smoothly as the simulation
progresses (runs A50, B50, B31 and B20). For the cases in which, as
stated above, there is practically no such structure (A31 and A20),
the accretion rate is episodic and directly reflects the periastron
passages of the surviving core. This is not easily seen at our current
level of resolution, since in these runs, the amount of mass
transferred in these events is small ($\Delta M\simeq 10^{-3}$) and
hence very few SPH particles are actually lost to the black hole (on
the order of 50). In Figure~\ref{mdot}a the mass transfer episode
occurring at $t\simeq 140$ is barely visible.

During our simulations, the total angular momentum in the system, {\em
J}, is not conserved for two reasons. First, since we have included a
gravitational radiation reaction term in the equations of motion, some
of it will be lost from the system.
\begin{figure*}
\psfig{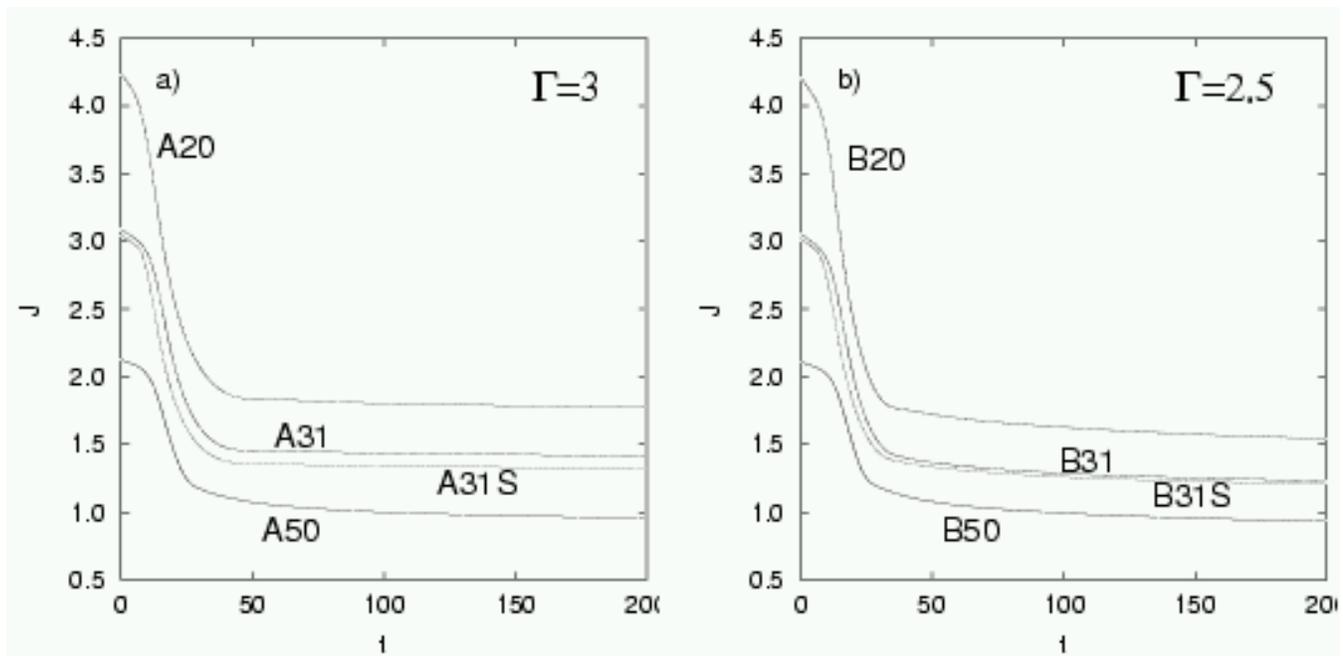}
\caption{Total angular momentum as a function of time for (a)
$\Gamma=3$ and (b) $\Gamma=2.5$ in units of $4.37\times
10^{42}$~kg~m$^{2}$~s$^{-1}$.}
\label{Jtot}
\end{figure*}
We have plotted {\em J} as a function of time in Figure~\ref{Jtot} for
all runs, and this effect can be seen in the early stages of the
simulation, before substantial mass transfer has taken place and
hydrodynamic effects have become important (before $t\simeq
10$). After this, the decrease is due primarily to accretion onto the
black hole (when matter is accreted through the horizon, the total
mass and linear momentum in the system are conserved, but not the
total angular momentum),
\begin{figure*}
\psfig{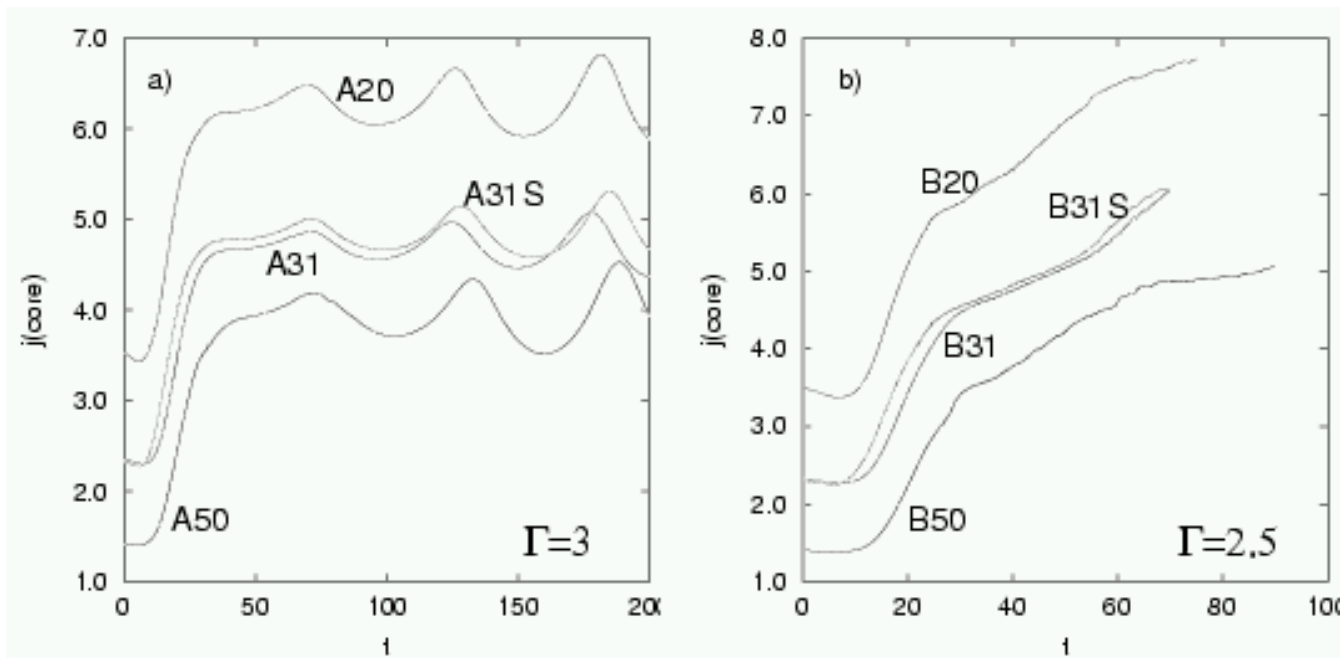}
\caption{Specific angular momentum of the core as a function of time
for (a) $\Gamma=3$ and (b) $\Gamma=2.5$ in units of $1.57\times
10^{12}$~m$^{2}$~s$^{-1}$. The curves in panel (b) terminate when the
core is disrupted.}
\label{jcore}
\end{figure*}
and we assume that it goes into spinning it up, thus giving it a
finite Kerr parameter $a=J_{\rm BH}c/G M_{\rm BH}^{2}$, shown in
Table~\ref{disks} (column 9) at the end of the simulation and assuming
that at $t=0$, $a=0$.

Why the neutron star core moves out after the first mass transfer
episode can be seen directly in Figure~\ref{jcore}, where we plot the
specific angular momentum of the core as a function of time. This
initially decreases slightly (for the same reasons we have outlined
above), and then increases sharply. The oscillations in the curves for
$\Gamma=3$ at late times reflect each successive periastron passage
and corresponding episode of mass transfer, and show that there is no
significant orbital decay (as is also evident from
Figure~\ref{rt}a). The slight variations arise because our numerical
definition of the core (for gravitational radiation reaction purposes)
makes its mass vary somewhat as the binary separation changes, thus
altering its total angular momentum content as well. The mass transfer
is not stable, and the evolution of the system at this stage is not
driven by the loss of angular momentum to gravitational waves as one
might initially expect but by hydrodynamic processes. This is the same
qualitative result that was obtained for tidally locked binaries with
a stiff equation of state (Klu\'{z}niak \& Lee~1998; Paper~I), and the
reason is essentially the same. For polytropes, the mass-radius
relationship is $R\propto M^{(\Gamma-2)/(3\Gamma-4)}$. For $\Gamma=3$
this becomes $R\propto M^{1/5}$ and for $\Gamma=2.5$, $R\propto
M^{1/7}$. Thus the neutron stars presented here respond to mass loss
by shrinking slightly, rather than expanding as should be the case
realistically (Arnett \& Bowers~1977). In the event of conservative
mass transfer in a binary system, if the donor is the lower mass
component, the orbital separation will increase. Strictly speaking we
are not dealing with a conservative system in this case, but the
general response of the system is the same. Since the neutron star
shrinks upon losing mass, it eventually cuts off the mass transfer
stream to the black hole when it no longer fills its Roche lobe, and
we are left with a stable binary. The surviving orbit is slightly
eccentric, and thus successive periastron passages allow a small
amount of mass transfer to take place. The evolution of the binary is
driven by two competing effects. On the one hand, gravitational
radiation reaction drains angular momentum from the system and thus
tends to make the separation decrease (it will also circularize the
orbit, but on a longer timescale than we have modeled here). On the
other, every time mass transfer occurs, it tends to increase the
separation as outlined above. It would appear from our calculations
that these effects cancel each other out on a timescale of many
orbits, but not in a manner that allows {\em steady} mass transfer, as
it has been suggested before (Blinnikov et al.~1984; Portegies
Zwart~1998). This is simply because the initial episode of mass
transfer (due to angular momentum loss to gravitational radiation, or
to a hydrodynamic instability) is much too violent to allow the star
to react fast enough. For softer equations of state with $\Gamma=2$
and $\Gamma=5/3$, whether the initial condition is that of a tidally
locked (Paper~II) or an irrotational binary (Lee~2000, in
preparation), the mass transfer process itself can be unstable and
lead to complete tidal disruption on a dynamical timescale (in this
case the mass--radius relationship is such that the star expands upon
losing mass). That stable mass transfer in a black hole--neutron star
binary is not possible had been pointed out by Bildsten \&
Cutler~(1992) and Kochanek~(1992) using essentially the same arguments
we have outlined above to explain our numerical results.

\begin{figure*}
\psfig{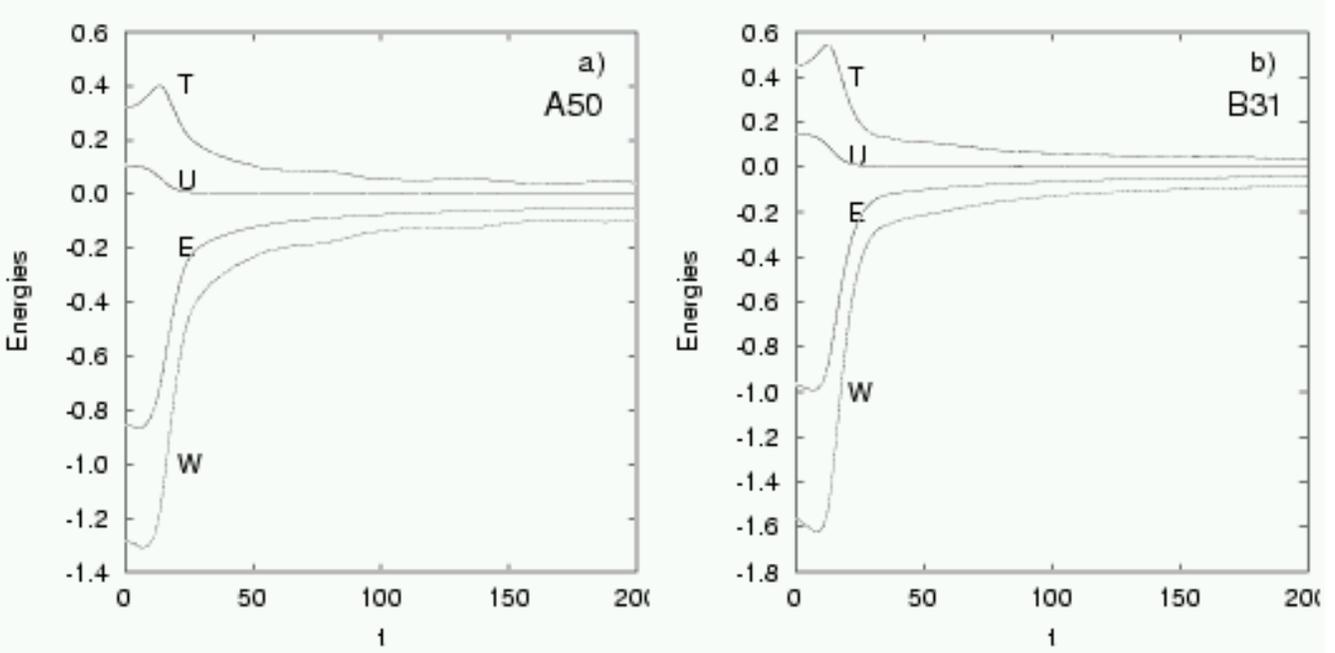}
\caption{Various energies in the system as a function of time for (a)
run A50 and (b) run B31. The kinetic (T), internal (U), gravitational
potential (W) and total (E) energies are indicated in units of
$3.8\times 10^{53}$~erg.}
\label{energies}
\end{figure*}

The various energies in the system are shown in Figure~\ref{energies}
for runs A50 and B31. The violent episode of mass transfer in the
initial stages of the coalescence is reflected in the drop in total
internal energy, while the survival of the binary in a slightly
eccentric orbit can be seen from the small oscillations in kinetic and
gravitational potential energy for run A50. In contrast, the
corresponding plot for run B31 shows only a monotonic change in the
curves at late times, when the neutron star has been completely
disrupted.

\subsection{Accretion disc structure}

In Table~\ref{disks} we show several parameters pertaining to the
final configuration of the system (at $t=t_{f}$). We calculate the
disc mass $M_{disc}$ by computing the amount of matter which has
angular momentum larger than a critical value, necessary for remaining
in orbit about the black hole. This matter has
$j>j_{crit}=\sqrt{6}GM_{total}/c^{2}$ (as in Paper~II). This value is
not zero even for the cases in which no appreciable disc has formed
(runs A31, A31S and A20 in particular) because we are {\em not}
excluding matter in the surviving core in this calculation. We are
assuming that the neutron star core will eventually be disrupted and
part of it will be accreted by the black hole. Of course for these
cases this is a very rough estimate since an explosion of the core
(see below) could redistribute angular momentum significantly. In the
cases where the neutron star is disrupted and a disc has formed, its
structure is becoming roughly azimuthally symmetric by the end of the
simulation (excluding the presence of the long tidal tail). Based on
the disc masses and final accretion rates (admittedly noisy and poorly
resolved) we infer a rough estimate for the lifetime of the disc as
$\tau_{disc}\simeq \dot{M}_{final}/M_{disc}$. The values shown in the
table are then between 70~ms and 100~ms. The viscosity in our
numerical scheme is purely artificial, so these values must be taken
with caution, and probably underestimate the true lifetime of the
disc.

\begin{figure*}
\psfig{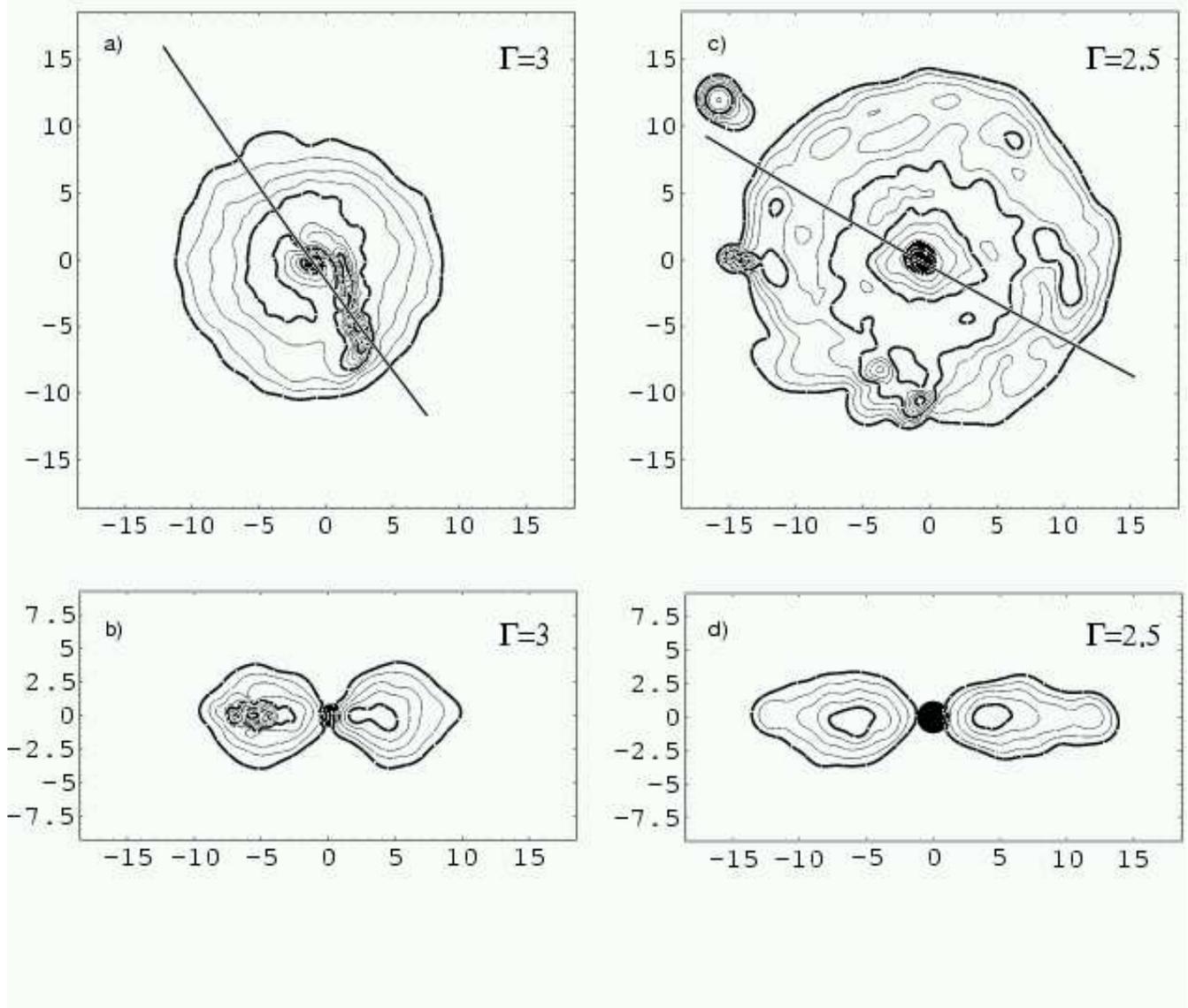}
\caption{Density contour plots at $t=t_{f}$ for runs A50~(a,b) and
B31~(c,d) in: (a,c) the orbital plane; (b,d) in the meridional plane
shown by the black line in panels (a,c). All contours are logarithmic
and equally spaced every 0.25 dex. Bold contours are plotted at $\log
\rho=-5,-4,-3,-2,-1$ (if present) in the units defined in
eq.~\ref{eq:defrhounit}.}
\label{rhog70}
\end{figure*}

\begin{figure*}
\psfig{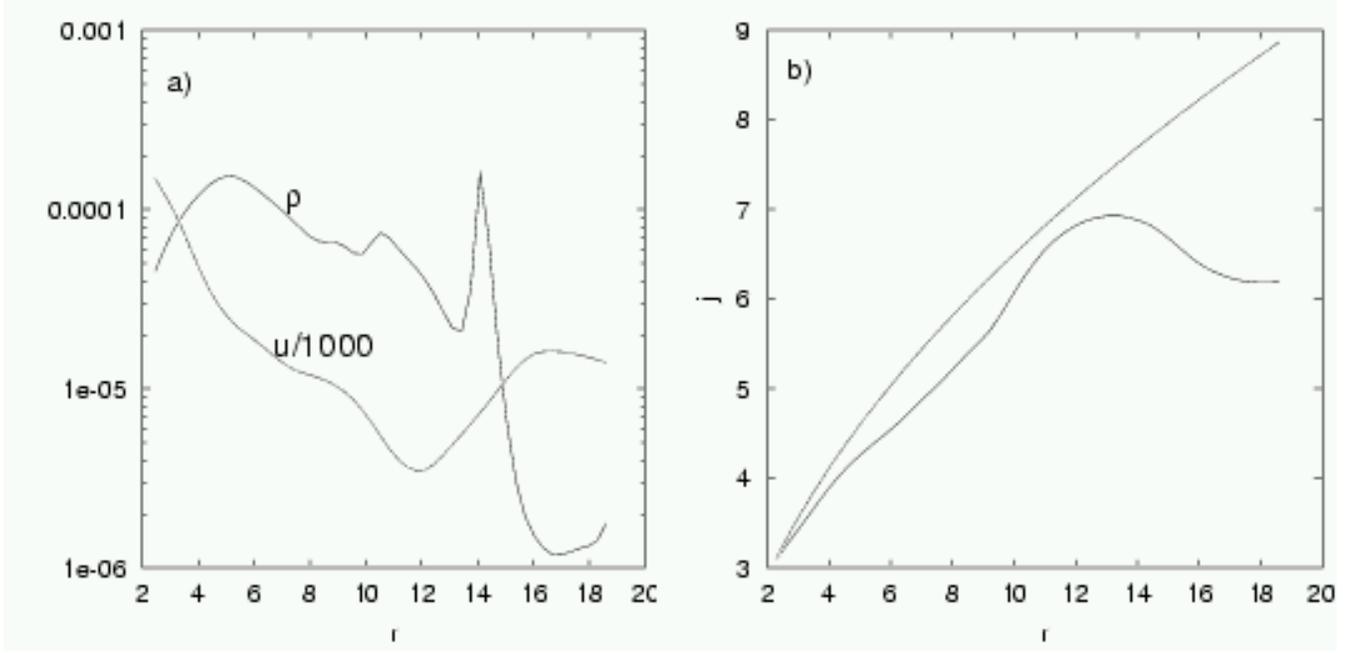}
\caption{Azimuthally averaged profiles for run B31 in the equatorial
plane for (a) the density $\rho$ and the specific internal energy $u$
($u/1000$ is plotted) and (b) the specific angular momentum (solid
line). The dashed line in (b) represents the rotation curve for a
Keplerian accretion disc around a black hole of the same
mass. Fluctuations in the curves reflect the fact that the disc is not
yet azimuthally symmetric. All curves terminate at $r=2r_{Sch}$.}
\label{diskprof}
\end{figure*}

We show in Figure~\ref{rhog70} density contours for runs A50 and B31
at the end of the simulations. 
In run A50, a secondary episode of mass transfer can be seen directly,
while a few of the knots formed in the initial tidal tail for run B31
can also be seen. The azimuthally averaged density and internal energy
profiles at $t=t_{f}$ for run B31 are shown in Figure~\ref{diskprof}.
The internal energy has a maximum in the inner regions of the disc,
and becomes somewhat flatter in the outer regions, at $u/1000\simeq
10^{-5}$, equivalent to $1.37\times 10^{18}$~erg~g$^{-1}$, or
1.45~MeV/nucleon. The fluctuations seen in the curves reflect the fact
that the accretion disc is not yet azimuthally symmetric. The rotation
curve is sub--Keplerian, indicating that pressure support is an
important factor in the structure of the disc.

\begin{figure*}
\psfig{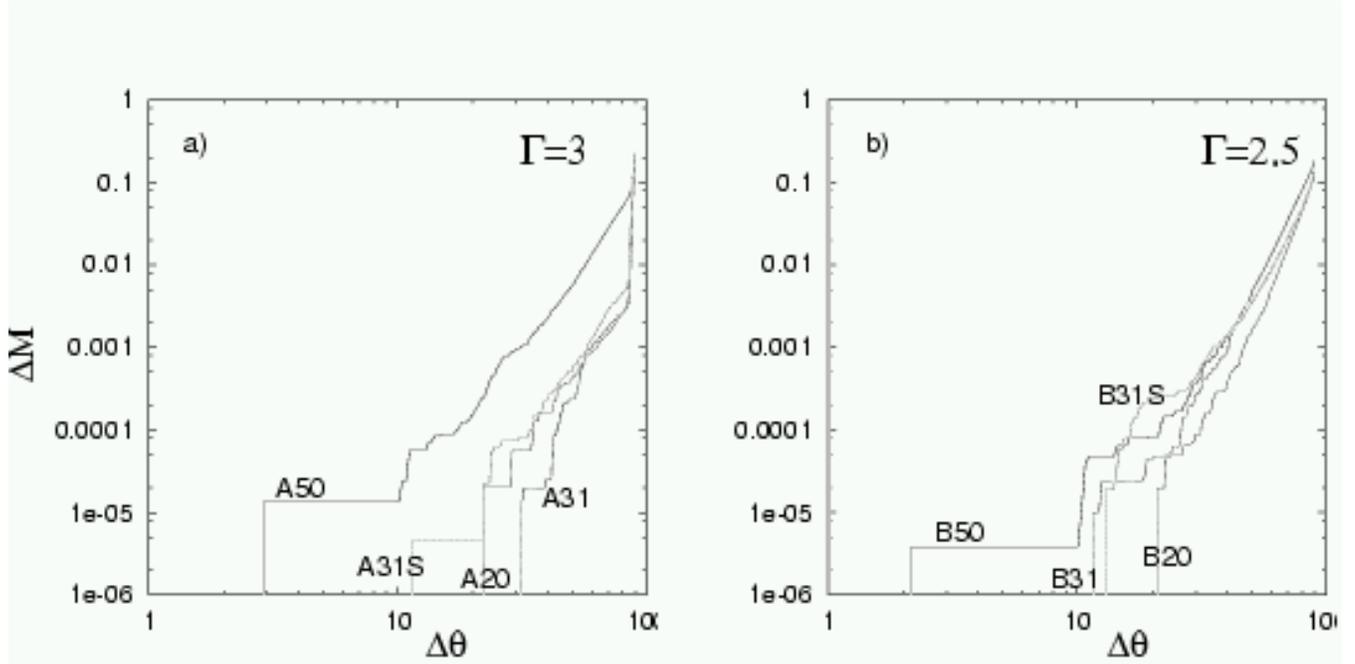}
\caption{Enclosed mass for all runs as a function of half--angle
$\Delta \theta$ (measured from the rotation axis in degrees) for (a)
$\Gamma=3$ and (b) $\Gamma=2.5$ at $t=t_{f}$. The sharp increase in
the curves in (a) for $80^{\circ} \leq \Delta \theta \leq 90^{\circ}$
shows the presence of the surviving core.}
\label{mtheta}
\end{figure*}

It is apparent in Figure~\ref{rhog70} that the region directly above
and below the black hole is largely free of matter. This is of crucial
importance in the context of gamma--ray bursts, since a low baryon
loading (of the order of 10$^{-5}$~M$_{\odot}$) is necessary to permit
the formation and expansion of an ultrarelativistic fireball
(M\'{e}sz\'{a}ros \& Rees~1992, 1993). Figure~\ref{mtheta} shows the
mass enclosed in a cone of opening angle $\Delta\theta$ directly above
and below the black hole, along the rotation axis. All runs show that
this axis is clear of matter within about 10$^{\circ}$ down to
approximately $10^{-5}$ M$_{\odot}$, and only modest collimation of an
outflow along the rotation axis is required to avoid contamination
(see the last three columns in Table~\ref{disks}). This is the first
time we have been able to reach such a low mass resolution limit
regarding the baryon pollution, and is due to the improved resolution
of our dynamical simulations (the runs shown in Papers~I,II used at
most $\simeq 17,000$ SPH particles). In the event of disc formation, a
fraction of the energy of rotation of the black hole could be
extracted from the system through the Blandford--Znajek~(1977)
mechanism. The black hole is spun up during the coalescence because of
the angular momentum of the accreted matter, to rates varying between
20\% and 40\% of the maximal rotation rate (see Table~\ref{disks}).

\begin{table*}
 \caption{Accretion disc structure}
 \label{disks}
 \begin{tabular}{@{}lccccccccccc}
  Run & $M_{disc}$ & $\dot{M}_{max}$    
	& $\dot{M}_{final}$ & $q_{final}$ & $M{\rm (core)}_{final}$
        & $M_{ejected}$
        & $\tau_{disc}$ & $J_{\rm BH}c/G M_{\rm BH}^{2}$ &
        $\theta_{-3}$
        & $\theta_{-4}$ & $\theta_{-5}$ \\
  A50   & 0.176 & 0.052 & $2\cdot 10^{-4}$ & 0.0286 & 
	0.080 & $2.908\cdot 10^{-2}$ & ...
        & 0.387 & 30 & 17 & 3 \\
  A31   & 0.182 & 0.051 & $1\cdot 10^{-4}$ & 0.0571 &
	0.227 & $2.238\cdot 10^{-2}$ & ...
        & 0.272 & 60 & 44 & 32 \\
  A31S  & 0.196 & 0.055 & $5\cdot 10^{-4}$ & 0.0462 &
	0.185 & $2.432\cdot 10^{-2}$ & ...
        & 0.277 & 58 & 35 & 22 \\
  A20   & 0.154 & 0.048 & $1\cdot 10^{-4}$ & 0.0319 &
	0.183 & $3.040\cdot 10^{-2}$ & ...
        & 0.190 & 63 & 35 & 22 \\
  B50   & 0.182 & 0.050 & $3\cdot 10^{-4}$ & ... &
	... & $2.377\cdot 10^{-2}$ & 607
        & 0.392 & 34 & 16 & 10 \\
  B31   & 0.170 & 0.053 & $2\cdot 10^{-4}$ & ... &
	... & $3.295\cdot 10^{-2}$ & 850
        & 0.288 & 38 & 26 & 12 \\
  B31S  & 0.167 & 0.057 & $2\cdot 10^{-4}$ & ... &
	... & $3.189\cdot 10^{-2}$ & 835
        & 0.286 & 33 & 16 & 13 \\
  B20   & 0.159 & 0.056 & $2\cdot 10^{-4}$ & ... &
	... & $3.156\cdot 10^{-2}$ & 795
        & 0.202 & 45 & 30 & 21 \\
 \end{tabular}

 \medskip

In the last three columns, $\theta_{-n}$ is the half--angle of a cone
above the black hole and along the rotation axis of the binary that
contains a mass $M=10^{-n}$.

\end{table*}

\subsection{Ejected mass and r--process}

\begin{figure}
\psfig{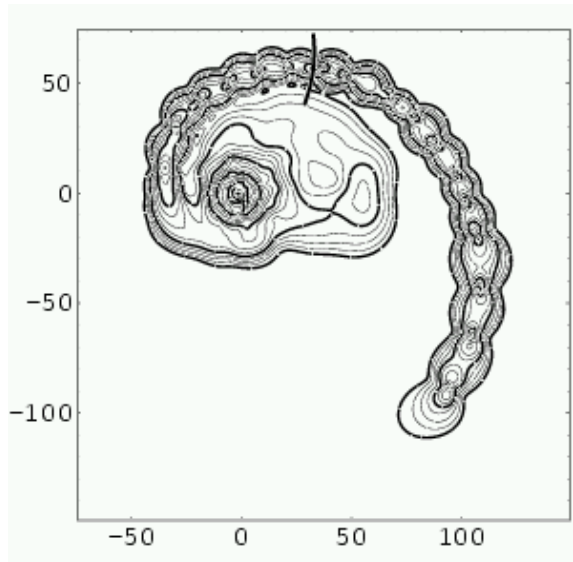}
\caption{Density contours in the orbital plane at $t=t_{f}$ for run
A50. All contours are logarithmic and equally spaced every 0.25
dex. Bold contours are plotted at $\log \rho=-8,-7,-6,-5,-4$ in the
units defined in eq.~\ref{eq:defrhounit}. The thick black line across
the tidal tail divides the matter that is bound to the black hole from
that which is on outbound trajectories. Only the fluid in the outer
tidal tail was considered when calculating the amount of ejected mass
shown in Table~\ref{disks}. The knots in the tail are clearly
visible.}
\label{tailsg3}
\end{figure}

We calculate the amount of mass that may be dynamically ejected from
the system as in Paper~II, by identifying matter with a positive total
energy (kinetic+gravitational potential+internal) at the end of each
simulation (see Table~\ref{disks}). This matter can be found in the
outer parts of the tidal tails that are formed during the initial
disruption of the star (for $\Gamma=2.5$) and also in the secondary
tails formed at secondary periastron passages of the core (for
$\Gamma=3$, see Figure~\ref{tailsg3}).
Since it is not clear that the fluid in this later category will truly
become unbound from the system (because of possible interactions with
other tidal tails or with the surviving core itself), we have not
included it in what we consider to be ejected mass, shown in
Table~\ref{disks} (this eliminates only about 10\% of the ejected mass
for $\Gamma=3$). Within our range of initial mass ratios, there
appears to be little difference in the amount of ejected mass. This
contrasts with the results of Janka et al.~(1999), who find that the
dynamical ejection of matter is highly suppressed at higher mass
ratios. The reason for this discrepancy could be due to their use of
the equation of state of Lattimer and Swesty, and to their different
implementation of gravitational radiation reaction (they use the
formalism of Blanchet, Damour \& Sch\"{a}fer~(1990) as opposed to our
use of the quadrupole approximation). We do, however, observe a slight
increase for the softer equation of state, probably because in this
case the star is completely disrupted, whereas for $\Gamma=3$ a
substantial amount of mass remains locked in the core. We find that
between $3.1\times 10^{-2}$ and $4.6\times 10^{-2}$~M$_{\odot}$ can be
dynamically ejected from the system. This is about the same as what
was found for double neutron star mergers by Rosswog et al.~(1999a),
who performed a Newtonian study and used the physical equation of
state of Lattimer \& Swesty (1991). A more detailed hydrodynamic study
has been carried out by Freiburghaus et al.~(1999b) to more accurately
determine the production of heavy elements through the r--process in
this ejected material. While our equation of state does not allow us
to follow such evolution, determining how much material can be
expelled to the interstellar medium from a dynamical coalescence is a
necessary first step. In this aspect we consider our results to be
upper bounds, mainly because the effects of general relativity will
probably make the ejection of neutron star material from the potential
well of the binary more difficult.

\subsection{Emission of gravitational waves}

\begin{figure*}
\psfig{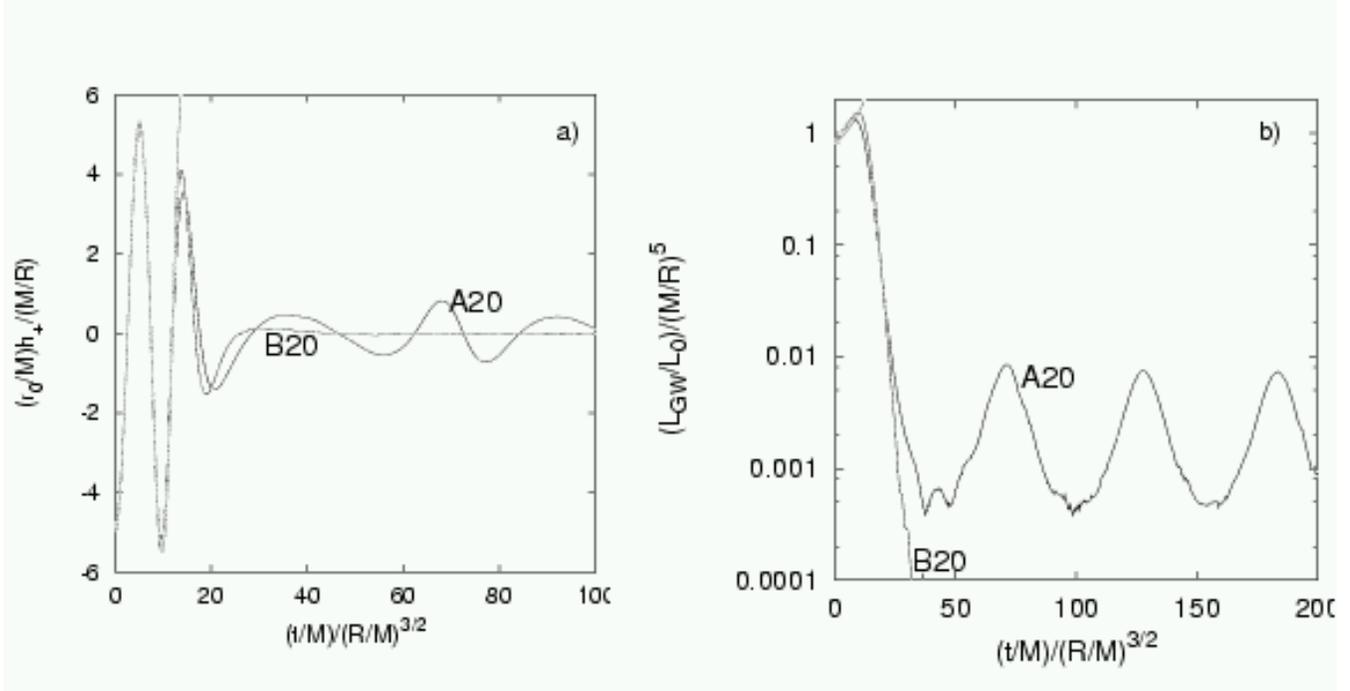}
\caption{(a) Gravitational radiation waveforms (one polarization is
shown) seen at a distance $r_{0}$ away from the system along the
rotation axis for runs A20 and B20. (b) Gravitational radiation
luminosity for the same runs as shown in (a). In both panels, the
dotted lines show the corresponding curves for a point--mass binary
with the same initial mass ratio and separation, decaying in the
quadrupole approximation. The survival of the binary in run A20 and
the tidal disruption in run B20 is apparent in both panels. All
quantitites are given in geometrized units such that $G=c=1$
($L_{0}=c^{5}/G=3.64\times 10^{59}$~erg~s$^{-1}$). We only plot the
curves for $0\leq t \leq 100$ in (a) to show detail during the initial
mass transfer event. Little further evolution occurs at later times.}
\label{gw}
\end{figure*}

As in previous work (see also Finn~1989; RS92) we have calculated the
emitted gravitational radiation waveforms during the merger in the
quadrupole approximation for every run. We show in Figure~\ref{gw} the
waveforms and luminosities for runs A20 and B20 (the curves are
qualitatively similar for runs A50, A31 and B50, B31 respectively),
compared to what these results would be for a point--mass binary
decaying through angular momentum loss in the quadrupole
approximation. Since the neutron star is completely disrupted for the
runs with $\Gamma=2.5$, the amplitude of the waveforms quickly drops
effectively to zero (we show upper bounds in Table~\ref{waves}), as
the accretion torus becomes ever more azimuthally symmetric. The
luminosity initially rises as the separation decreases (before any
significant amount of mass transfer has taken place, until $t\simeq
10$) and then decays quickly, exhibiting a single, sharp peak. For
$\Gamma=3$ the behavior is initially similar, except that now there
remains a coherent signal that clearly has the signature of a binary
system, since the core remains in orbit around the black hole for the
duration of the simulation. The orbit is slightly elliptical, and the
successive peaks in the gravitational radiation luminosity (at
$t\simeq 75, 130, 185$) correspond to the later periastron passages of
the core. We detail the information extracted from the waveforms in
Table~\ref{waves}, where we show the maximum and final amplitudes for
the waveforms, the peak luminosity and the total energy radiated away
by the system, and the efficiency of gravitational wave emission
$\epsilon=\Delta E/M_{\rm total}c^{2}$. For reference, $L_{max}=1$ (in
the units given in the table) corresponds to $3.036\times
10^{55}$~erg~s$^{-1}$and $\Delta E=10$ is equivalent to $3.48\times
10^{52}$~erg. The variations for different values of the mass ratio
(at fixed $\Gamma$) arise because we have normalized results to the
initial neutron star mass (1.4~M$_{\odot}$). However, it is clear that
there are quantitative differences depending on the adiabatic index
(at a fixed initial mass ratio). The stiffer the equation of state,
the more important tidal effects are, and these contribute to orbital
decay, as we have seen. Thus the binaries with $\Gamma=3$ are
disrupted earlier during the encounter, reaching consistently lower
values in amplitude for $h_{max}$ and $L_{max}$. The total energy
emitted by the system should not be taken as an absolute value, since
it depends on the choice of the origin of time. We can nevertheless
perform comparisons between runs with this figure, since our initial
separations are always comparable for a given choice of $q$. Here too
we see that less energy is emitted for the stiffer equation of state,
despite the fact that the binary survives and continues to produce
gravitational waves. This is simply because the separation and mass
ratio have changed so much during the mass transfer episodes that
emission is quickly suppressed by more than two orders of magnitude
(see Figure~\ref{gw}b).

\begin{table*}
 \caption{Gravitational radiation}
 \label{waves}
 \begin{tabular}{@{}lccccc}
  Run & $(r_{0}R/M_{\rm NS}^{2})h_{max}$
   	& $(r_{0}R/M_{\rm NS}^{2})h_{final}$
        & $(R/M_{\rm NS})^{5}(L_{max}/L_{0})$
        & $(R^{7/2}/M_{\rm NS}^{9/2})\Delta E_{GW}$ & $\epsilon$ \\
      &      &      &      &      \\
  A50   & 2.80 & 0.20 & 0.42 & 5.58  & $2.59\cdot10^{-3}$ \\ 
  A31   & 3.95 & 0.70 & 0.78 & 10.52 & $3.47\cdot10^{-3}$ \\
  A31S  & 3.99 & 0.50 & 0.83 & 8.84  & $2.92\cdot10^{-3}$ \\ 
  A20   & 5.37 & 0.80 & 1.34 & 16.59 & $3.85\cdot10^{-3}$ \\ 
  B50   & 2.88 & $\leq$ 0.01 & 0.48 & 6.26  & $2.91\cdot10^{-3}$ \\ 
  B31   & 4.02 & $\leq$ 0.02 & 0.87 & 11.03 & $3.64\cdot10^{-3}$ \\ 
  B31S  & 4.06 & $\leq$ 0.02 & 0.93 & 9.49  & $3.13\cdot10^{-3}$ \\ 
  B20   & 5.46 & $\leq$ 0.02 & 1.54 & 18.30 & $4.25\cdot10^{-3}$ \\ 

 \end{tabular}

 \medskip 

All quantitites are given in geometrized units such that $G=c=1$, and
$L_{0}=c^{5}/G=3.64\times 10^{59}$~erg~s$^{-1}$.
\end{table*}

\section{Influence of initial conditions on the dynamical evolution 
of the system} \label{sphere}

As mentioned before, we have performed two runs, A31S and B31S, in
which the initial condition uses a spherical neutron star, relaxed in
isolation. The initial orbital angular velocity used for these
dynamical runs is that of a point mass binary with the same initial
mass ratio and separation. Otherwise all initial parameters are as in
runs A31 and B31 respectively (see
Table~\ref{parameters}). Constructing fully self--consistent initial
conditions for close binaries is not an easy task, particularly for
non--synchronized binaries like the ones treated in this paper. The
complications that arise also depend on the particular numerical
method being used, and there is a trade--off between numerical
accuracy and physical sense of the results that is always present. All
other things being equal, at larger binary separations, tidal effects
are smaller (how much smaller depends on the details of the equation
of state), and thus taking a spherical star as an initial condition
can be quite reasonable. However, one then needs to evolve the system
for a large number of dynamical times (e.g. draining angular momentum
through gravitational wave emisison) before the actual coalescence
takes place. This has two very important drawbacks. First, much CPU
time is wasted modeling the system at high resolution while the
hydrodynamics has little effect on the evolution. Zhuge et al.~(1994,
1996) have performed dynamical simulations of binary neutron star
coalescence using SPH, starting with initial conditions at large
separations and removing angular momentum from the system through
gravitational wave emission as we have done here. Their spatial
resolution was severely limited (the neutron star was modeled with
approximately 4000 particles at most) because they had to evolve the
system for a large number of dynamical times. Second, spurious
numerical effects can degrade the initial condition one imposed at the
start (through viscous dissipation and angular momentum transport) so
that we have little control over the physical state of the system when
the final merging occurs. Thus, many of the simulations of binary
coalescence reported in the literature for double neutron star systems
(e.g. Oohara \& Nakamura~1989; Benz et al.~1990; Davies et al.~1994,
Ruffert et al.~1996; Rosswog et al.~1999), and for black hole neutron
star binaries (Janka et al.~1999), have used spherical neutron stars
as an initial condition, regardless of the initial separation, which
is nevertheless small enough so that the two problems stated above are
avoided. We note here that it is much easier to build even approximate
equilibrium solutions to initial conditions if the equation of state
is of a simple form like the one we have used here, rather than a
physical equation of state like the one used by Janka et al.~(1999)
and Rosswog et al.~(1999) (this is another trade-off one must
balance). Here we wish to explore how one aspect (tri--axial ellipsoid
vs. spherical star) of the choice of initial conditions will affect
the outcome of the coalescence, in a qualitative and quantitative
fashion. As noted above (see section~\ref{initial}), we remind the
reader that an irrotational Roche--Riemann ellipsoid does not
correspond to a fully self--consistent initial condition either.

\begin{figure*}
\psfig{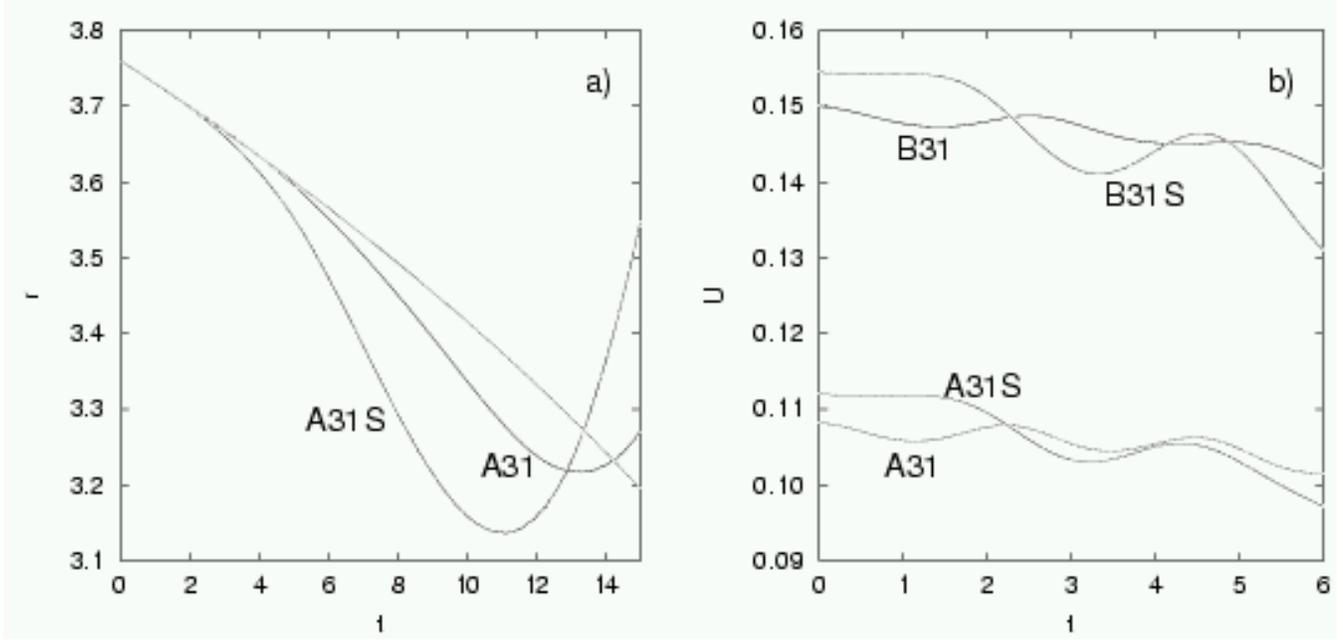}
\caption{(a) Binary separation as a function of time for runs A31 and
A31S. The monotonically decaying line corresponds to a point--mass
binary with the same initial mass ratio and separation decaying
through gravitational wave emission, computed in the quadrupole
approximation. The behavior is qualitatively similar for runs B31 and
B31S. (b) Total internal energy of the neutron star as a function of
time for runs A31, A31S, B31 and B31S. The oscillations arising from
the formation of the tidal bulge are clearly visible.}
\label{decay}
\end{figure*}

\begin{figure*}
\psfig{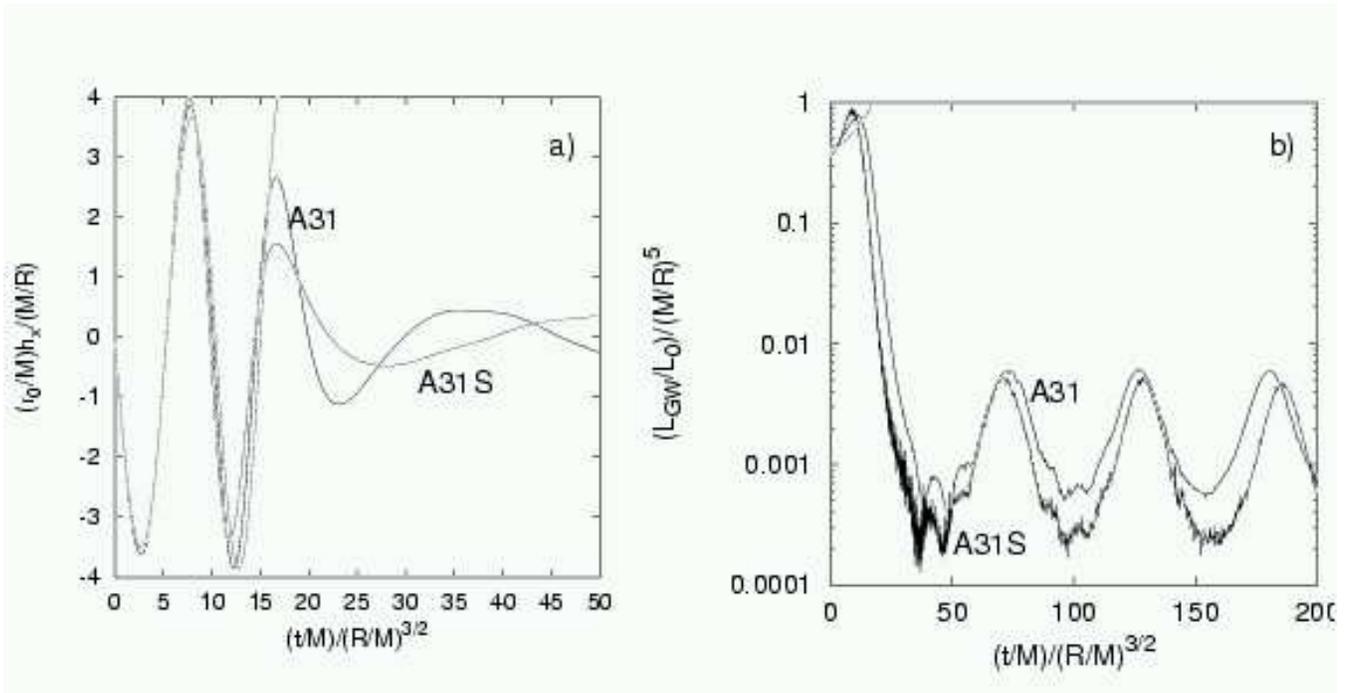}
\caption{(a) Gravitational radiation waveforms (one polarization is
shown) seen at a distance $r_{0}$ away from the system along the
rotation axis for runs A31 and A31S. (b) Gravitational radiation
luminosity for the same runs as shown in (a). In both panels, the
dotted lines show the corresponding curves for a point--mass binary
with the same initial mass ratio and separation, decaying in the
quadrupole approximation.  All quantitites are given in geometrized
units such that $G=c=1$ ($L_{0}=c^{5}/G=3.64\times
10^{59}$~erg~s$^{-1}$).}
\label{gwb}
\end{figure*}

The main effect of using a spherical star as an initial condition
appears as soon as the dynamical simulation starts, since the neutron
star must respond the the instantaneous appearance of a tidal field
produced by the black hole. Thus a tidal bulge quickly appears,
initially along the axis joining the two binary components. This act
drains energy from the orbital motion, and makes the separation decay
slightly faster than it otherwise would (see Figure~\ref{decay}a).
The appearance of the tidal bulge also induces radial oscillations of
the star which damp out after a few dynamical times, and can be
appreciated in Figure~\ref{decay}b, where we plot the total internal
energy of the neutron star for runs A31, A31S, B31 and B31S. All
curves show variation, since runs A31 and B31 do not correspond to
full equilibrium solutions either, but their oscillations are somewhat
smaller. This initial perturbation is responsible for all subsequent
differences in the variables shown in
Tables~\ref{disks}~and~\ref{waves}, and are easily understood. Since
the separation decreases at a higher rate, the initial encounter with
the black hole is more violent, making the peak accretion rates larger
by about 8\%, and the surviving core (for $\Gamma=3$) less massive by
about 20\%. The gravitational wave signal (see Figure~\ref{gwb})
is also affected for the same reasons. The peak amplitude in the
waveforms and the peak luminosity are slightly higher (because the
minimum separation reached during the first episode of mass transfer
is smaller) but decay faster, and the final amplitude (when the binary
survives) is smaller (largely because the final mass ratio is
lower). However, the total energy radiated away through gravitational
waves is significantly lower (by about 15\%), because the initial
encounter is much more brief. Finally, the faster initial decay is
also reflected in a slight phase shift in the gravitational wave
signal (see Figure~\ref{gwb}b).

\section{Discussion and conclusions} \label{discussion}

We have presented the results of three--dimensional hydrodynamical
simulations of the coalescence of a neutron star with a black hole. We
use a stiff equation of state to model the neutron star as a cold
polytrope with $\Gamma=3$ and $\Gamma=2.5$, which is believed to be
appropriate for matter at nuclear densities. The black hole is modeled
as a Newtonian point mass with an absorbing boundary at the
Schwarzschild radius. We have explored a range of mass ratios
$q=M_{\rm NS}/M_{\rm BH}$, and follow the dynamical evolution of the
system for approximately 23~ms. It is important to keep in mind that
our simulations are completely Newtonian, except for our treatment of
gravitational radiation reaction on the hydrodynamics (see
section~\ref{grr}). Since tidal locking is not expected in this type
of system (Bildsten \& Cutler~1992; Kochanek~1992) we have used
initial conditions corresponding to irrotational binaries in
equilibrium, approximating the neutron star as a compressible
tri--axial ellipsoid (see LRSb). The dynamical simulations are begun
when the system is on the verge of initiating mass transfer.

The loss of angular momentum to gravitational wave emission, together
with hydrodynamic instabilities, always drives these systems to
initiate mass transfer within one orbital period after the start of
the dynamical simulations. For $\Gamma=3$ this episode always leads to
the survival of the neutron star core and its transfer to a higher
orbit. A long tidal tail of matter stripped from the star appears from
the outer Lagrange point, and for high mass ratios ($q=0.5$ in our
simulations) an accretion disc is clearly visible (for lower mass
ratios this is not so evident at our current level of resolution). The
new binary system is clearly stable, and the slightly elliptical orbit
($0.1\leq e \leq 0.2$) is small enough to allow secondary mass
transfer events to occur at each periastron passage (two or three more
occur during our simulations). Since the mass ratio in the new binary
differs substantially from the initial one, and the separation is
larger, the timescale for decay of the orbit due to gravitational wave
emission is lenghtened considerably, and becomes up to two orders of
magnitude greater than the orbital period (calculated for point--mass
binaries in the quadrupole approximation). Thus the lifetime of the
system may be on the order of tenths of one second.  For $\Gamma=2.5$,
the neutron star is eventually completely disrupted and an accretion
disc is formed around the black hole, but over a longer timescale
(essentially up to and including the second periastron passage of the
core) than in the previous case. A tidal tail of material thrown out
from the outer Lagrange point is also formed as described above. These
tails survive as well--defined structures throughout the simulations.

The overall morphology of the events is directly reflected in the
gravitational radiation waveforms. In the cases where the stellar core
survives, the signal continues to exhibit a finite amplitude (albeit
greatly reduced) at a frequency corresponding to the new orbital
period (approximately 350~Hz vs. 800~Hz at the start of the dynamical
simulations, since the frequency of the gravitational radiation signal
is twice that of orbital motion for circular orbits). The eccentricity
of the orbit produces periodic peaks in the gravitational wave
luminosity. This remnant signal is completely absent in the case of
$\Gamma=2.5$, where the waveforms (and luminosity) drop apruptly and
practically to zero after the star is disrupted and the accretion disc
is formed. The dramatic dependence on the stiffness of the equation of
state shows the type of information that could be gleaned from an
observation of gravitational waves in the near future, and is similar
to what has been found for double neutron star systems by Rasio \&
Shapiro~(RS94), where the persistent emission was due to a lack of
axisymmetry in the central object left after the coalescence.

In all cases, the fluid contained in the outer parts of the tidal
tails (amounting to 10$^{-2}$--10$^{-1}$~M$_{\odot}$) appears to be on
outbound trajectories (it has enough mechanical energy to escape the
gravitational potential well of the black hole--debris torus
system). This probably represents an upper bound, since the effects of
general relativity are likely to lower this value. As we have pointed
out above, this may have important implications for the abundances of
r--process material in the galaxy. We refer the reader to the more
detailed analysis performed by Rosswog et al.~(1999) and Freiburghaus
et al.~(1999b) where they have used the equation of state of Lattimer
\& Swesty~(1991) and a detailed thermodynamic and nuclear network
calculation.

The use of adequate initial conditions for detailed dynamical
simulations is always an important concern. For tidally locked
binaries, setting up equilibrium initial conditions can be done in a
relatively simple way (Paper~I, Paper~II). This is not the case if
this restriction is lifted, as it has been for this work. We have
chosen to use the semi--analytical method of LRSb to construct
approximate equilibrium solutions to the problem of a compressible
tri--axial ellipsoid in orbit about a point mass companion (see
Ury\={u} \& Eriguchi~1999 for full equilibrium configurations). We
remark that this does not mean that our initial conditions are
entirely self--consistent, since at small separations, a dynamical
tidal lag angle is induced in the binary (see Lai et al.~1994), and we
have necessarily neglected this effect for now. We have carried out
two dynamical simulations in this work that use a spherical neutron
star relaxed in isolation as an initial condition, and find that there
are quantitative differences in the evolution of the system (see
section~\ref{sphere}). These can be directly traced to the fact that
the spherical star must respond instantaneously to the presence of the
gravitational field of the black hole when the dynamical simulation is
started. This leads to radial oscillations of the star and to a more
rapid orbital decay, as compared to the equilibrium case. The
evolution is more rapid and the mass transfer episodes are more
violent. Thus the maximum accretion rates and the peak in the
gravitational radiation waveforms and luminosities are overestimated
(see Table~\ref{waves} and Figure~\ref{gwb}).

We can perform a direct comparison between run A31 and the dynamical
coalescence presented in Paper~I for a tidally locked binary with
initial mass ratio $q=0.31$ (see section~5.4.2 and Figures~20--23 in
Paper~I). Both simulations included the effect of gravitational
radiation reaction on the system and were carried out with an
adiabatic index $\Gamma=3$, thus any differences are due entirely to
the initial condition. In both cases there is an initial episode of
intense mass transfer, with the surviving core being transferred to a
higher orbit. However, in the case of the tidally locked system, it is
less violent, and thus the final separation of the binary is smaller
than in run A31, and the orbit is more circular. This allows the
secondary episodes of mass transfer to be more intense, by about an
order of magnitude (they are clearly resolved in the run shown in
Paper~I, despite the lower numerical resolution). The evolution is
reflected in the gravitational radiation signal, with the final
amplitude of the waveforms being greater by a factor of two than in
run A31, and the gravitational radiation luminosity exhibiting
secondary peaks at each periastron passage that are greater by
approximately one order of magnitude. Additionally, no resolvable
amount of mass ($M_{ejected}\leq10^{-4}$) is dynamically ejected in the
tidally locked case, in stark contrast with the result of run A31.

The surviving core (for $\Gamma=3$) can be driven below the minimum
mass required for stability by the successive episodes of mass
transfer. If this occurs, an explosion may take place (Page~1982;
Blinnikov et al.~1984; Colpi, Shapiro \& Teukolsky~1991; Sumiyoshi et
al.~1998). Exploring this process clearly requires a greater level of
detail in the physics input than we have at our disposal for this
work, and is beyond the scope of this paper. The presence of the black
hole would make this even more complicated.

All the accretion discs that form around the black hole during the
coalescence are similar in structure, and by the end of our
simulations, they are quite close to being azimuthally symmetric. They
have masses of a few tenths of a solar mass, with maximum densities
and specific internal energies of order 10$^{11}$g~cm$^{-3}$ and
10$^{19}$erg~g$^{-1}$ (or 10~MeV/nucelon) respectively. All of the
final configurations have a low degree of baryon contamination along
the rotation axis, in the regions directly above and below the black
hole. It is low enough so that only modest beaming (of approximately
10$^{\circ}$) of a relativistic fireball along this axis would be
required in order to avoid being stopped by the fluid in the vicinity
of the black hole (M\'{e}sz\'{a}ros \& Rees~1992, 1993). This is
encouraging as far as current models of gamma ray bursts (GRBs) are
concerned, and confirms our previous results (Lee \&
Klu\'{z}niak~1997; Klu\'{z}niak \& Lee~1998), where we found that
these systems appear to be good candidates for the central engines of
short GRBs. The recent simulations of black hole--neutron star
coalescence performed by Janka et al.~(1999) using the equation of
state of Lattimer \& Swesty~(1991) have produced essentially the same
results. The coalescence of binary neutron stars may result in a very
similar scenario if the central object collapses to a black hole
(Ruffert \& Janka~1999).

\section*{Acknowledgments}

It is a pleasure to acknowledge many helpful discussions with W\l
odzimierz Klu\'{z}niak, Frederic Rasio, Lars Bildsten and Maximilian
Ruffert. I thank the Aspen Center for Physics for its
hospitatlity. Support for this work was provided by CONACyT (27987E)
and DGAPA--UNAM (IN-119998). I thank the anonymous referee for his
comments and suggestions for improvements to the text.

\label{lastpage}

\end{document}